\documentclass[prb,reprint]{revtex4-2}
\usepackage{graphicx}
\usepackage{dcolumn}
\usepackage{bm}
\usepackage[svgnames]{xcolor}
\usepackage{mathtools}
\usepackage{dsfont}
\usepackage{url}
\usepackage{hyperref}
\usepackage[capitalise]{cleveref}
\usepackage{braket}
\usepackage[version=4]{mhchem}
\usepackage{ragged2e}
\usepackage{subcaption}
\usepackage{comment}
\usepackage{amssymb}
\usepackage{mathrsfs}
\usepackage{xcolor}

\usepackage{ulem}
\usepackage{soul}

\begin{document}

\preprint{APS/123-QED}

\title{Topological Amplification of the Bosonic Kitaev Chain with Non-Uniform Loss}

\author{Cl\'ement Fortin$^1$}
\author{Kai Wang$^{1}$}
\author{T. Pereg-Barnea$^{1,2}$}

\affiliation{$^1$Department of Physics and Regroupement qu\'{e}b\'{e}cois sur les mat\'{e}riaux de pointe, McGill University, Montr\'eal, Qu\'ebec H3A 2T8, Canada \\
$^2$ICFO-Institut de Ciencies Fotoniques, The Barcelona Institute of
Science and Technology, Castelldefels (Barcelona) 08860, Spain}

\date{\today}

\begin{abstract}
The bosonic Kitaev chain is known to have extraordinary properties distinct from its fermionic counterpart. For example, it exhibits the non-Hermitian skin effect---its eigenmodes are exponentially localized to the edges of the chain---even when the system is Hermitian. Such non-Hermitian effects originate from the fact that the dynamics of bosonic quadratic Hamiltonians is governed by a non-Hermitian matrix. In the topological phase of the model, the modes conspire to lead to phase-dependent and directional exponential amplification of a classical drive. In this work, we study the robustness of this topological amplification to on-site dissipations. We examine the effect of uniform and non-uniform losses under various configurations. We find a remarkable resilience to dissipation in some configurations, while in others the dissipation causes a topological phase transition which eliminates the exponential amplification. In particular, when the dissipation is placed on every other site, the system remains topological and the exponential amplification persists even for very large loss rates which exceed the system's non-Hermitian gap. On the other hand, we find that dividing the chain into unit cells of an odd number of sites and placing dissipation on the first site leads to a topological phase transition at a certain critical value of the dissipation. Our work thus provides insights into the robustness against losses of the topological amplification of non-Hermitian systems and sets explicit limits on the bosonic Kitaev chain's ability to act as a multimode quantum sensor in realistic lossy scenarios.

\end{abstract}

\maketitle

\section{Introduction}\label{sec:Intro}
Physical systems exhibiting nontrivial topological invariants constitute a new path for quantum sensing protocols, where one requires the sensor to display high sensitivity to specific perturbations while being robust to other common perturbations. Bosonic quadratic systems are of particular interest for such applications as they manifest nontrivial topology and exotic effects of highly sensitive amplifications \cite{McDonald2018,Porras2019,Wanjura2020,McDonald2020,Ramos2021,Gomez2022}. These bosonic models differ fundamentally from their fermionic counterparts, in that a Hermitian Hamiltonian can give rise to non-Hermitian (NH) dynamics \cite{Lieu2018,Wang2019,Flynn2020}. They are also standard platforms for studying magnons \cite{Katsura2010,Kim2016} and the quadrature squeezing of light \cite{Walls1983,Caves1985}---notably used to enhance gravitational wave sensitivity \cite{Caves1981,Ligo2013,Virgo2019} and perform quantum computations \cite{Braunstein2005}. 

Features unique to NH systems, such as exceptional points \cite{Wiersig2014,Wiersig2016,Wiersig2020,Liu2016,Hodaei2017,Chen2017,Zhang2019,El-Ganainy2018,Ren2017,Wang2024,Luo2022,Lai2019} and the non-Hermitian skin effect \cite{Yao2018, Martinez2018, Budich2020,Ehrhardt2024,Sarkar2024,Bao2022} have recently been proposed as resources for enhanced sensitivity in the context of quantum sensing.
Perhaps most promising is nonreciprocal transport \cite{McDonald2020}, which is tightly linked to directional amplifiers \cite{Deak2012,Caloz2018} and has the advantage of not requiring fine parameter tuning (as is the case for exceptional-point-based sensing) or high intracavity photon number and reflection gain \cite{Lau2018}. 
It can be realized via reservoir engineering \cite{Metelmann2015,Metelmann2017,Fang2017}, Josephson circuits \cite{Kamal2011,Abdo2013,Sliwa2015,Ramos2024} and optomechanical systems \cite{Malz2018,Shen2016,Ruesink2016,Xu2016,Bernier2017,Peterson2017}. 

In the bosonic Kitaev chain (BKC) \cite{McDonald2018,Kitaev2001}, nonreciprocity caused by nearest-neighbor squeezing leads the system to be highly sensitive to $\mathbb{Z}_2$ symmetry-breaking perturbations \cite{McDonald2020}. This sensitivity manifests itself through an exponential scaling in system size of the steady-state susceptibility, or Green's function. The particularity of the system causes the signal and noise due to such perturbations to scale differently, leading to an exponentially-large signal-to-noise ratio \cite{McDonald2020}. Some of these phenomena have recently been observed in optomechanical networks \cite{Slim2024} and superconducting circuits \cite{Busnaina2024}.

\begin{figure}[b!]
    \centering
    \includegraphics[width=0.47\textwidth]{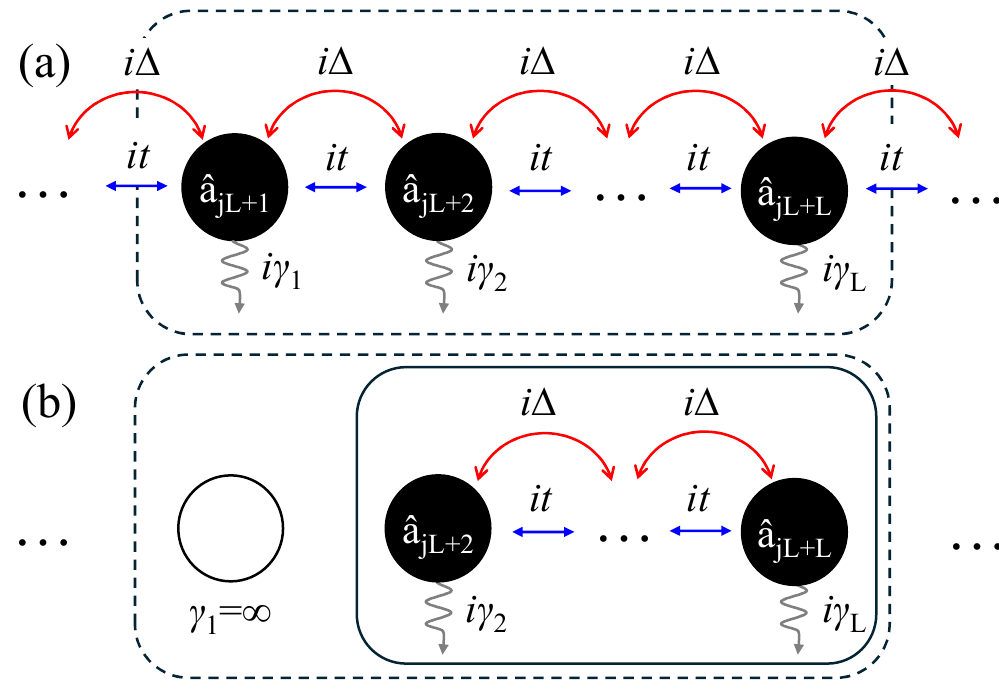}
    \caption{\justifying (a) Schematic of a unit cell (dashed box) of length $L$ for the BKC subject to on-site dissipation $\gamma_j\geq 0$. (b) When $\gamma_1\to\infty$, the unit cell effectively becomes a BKC with $L-1$ sites and OBC (solid box).}
    \label{fig:unit_cell}
\end{figure}

Remarkably, studies revealed that the exponential scaling of the susceptibility is a NH topological phenomenon \cite{Wanjura2020,Wanjura2022}. Indeed, the behavior of the susceptibility for one-dimensional driven-dissipative Hamiltonians under open boundary conditions (OBC) is associated with a topological invariant of the dynamical matrix under periodic boundary conditions (PBC). This leads to the so-called topological amplification \cite{Porras2019,Brunelli2023}. Interestingly, this phenomenon persists in dynamically metastable regimes, some of which can only be accessed through dissipation \cite{Flynn2020,Flynn2021,Gomez2023,Flynn2023,Ughrelidze2024}. 

Using the singular value decomposition, topological amplification provides a bulk-boundary correspondence \cite{Porras2019,Herviou2019,Brunelli2023} where the value of the topological invariant corresponds to the number of channels for amplification in the steady state \cite{Brunelli2023}. Such a NH invariant thus appears differently in long-time dynamics than eigenstate-based invariants \cite{Cardano2017,Maffei2018,Longhi2018_2,Jiao2021,Caceres2023,Villa2024}.

As with Hermitian systems, the topological phase in NH systems is preserved when the dissipation does not exceed the size of the relevant gap \cite{Wanjura2022}, here defined as the minimum distance from the PBC complex energy curve to the origin of the complex energy plane \cite{Brunelli2023}. We find that this criterion, while sufficient, is not a necessary one. While previous studies on the impact of loss for the BKC in this context focused on uniform dissipation \cite{Porras2019,Wanjura2020,Ramos2021,Gomez2022}, we find cases in which non-uniform dissipation can greatly exceed the above limit. These results thus indicate that a unified understanding of topological amplification in disordered systems remains to be established. Further, we note that utilizing these realizations of the BKC as quantum sensors and amplifiers would require the precise delineation of the role of non-uniform loss and this paper takes a step in this direction.

In this work, we study the limits of topological amplification in the bosonic Kitaev chain subject to non-uniform loss. We do this by dividing the system into unit cells of $L$ sites, see \cref{fig:unit_cell} (a).

In \cref{sec:Model}, we present the model in more detail and describe topological amplification using the singular value decomposition. We start by applying this framework to the case of uniform dissipation in \cref{sec:uniform}. We then look at how dissipation affects the PBC spectral curve of the dynamical matrix for different unit cell sizes in \cref{sec:band_separation} by studying the chain's symmetries. 
In \cref{sec:robutness}, we show that the conventional expectation on robustness to loss can be drastically exceeded: the bosonic Kitaev chain with an even number of sites always exhibits topological amplification provided that loss is placed on every other site. To understand why, we treat this situation in \cref{subsec:even_unit_cell} as a disordered version of the dissipative bosonic Kitaev chain whose unit cell has two sites and where only one is lossy. We find that dissipation induces a topological phase transition only when both sites of the unit cell are allowed to be lossy. Moreover, we show that the PBC energy band splits in two when the absolute difference of the two sites' bath coupling constant exceeds a critical value, but that topological amplification remains as long as one band winds around the origin. Lastly, we study robustness to dissipation in larger unit cells in \cref{subsec:general_L}. The code used to generate our data and plots can be accessed in Ref.~\cite{code2025}.

\section{The BKC model, input-output theory and topological classification}\label{sec:Model}


The bosonic Kitaev chain \cite{McDonald2018} is defined as 
\begin{align}
    \label{eq:BKC}
    \hat{\mathcal{H}} &= \sum_j \left(it\hat{a}_{j+1}^\dagger \hat{a}_j + i\Delta \hat{a}_{j+1}^\dagger \hat{a}_j^\dagger + h.c.\right)
\end{align}
where $\hat{a}_j^\dagger,\hat{a}_j$ are the bosonic creation and annihilation operators on site $j$, respectively, satisfying $[\hat{a}_i,\hat{a}_j^\dagger] = \delta_{i,j}$. We study the model for both periodic boundary conditions (PBC) and open boundary conditions (OBC). The nearest-neighbor hopping $t$ and squeezing strength $\Delta$ are assumed to satisfy $t>\Delta>0$ as this corresponds to the dynamically stable regime of the open-boundary system \cite{McDonald2018}. The remarkable features of the model are best understood by considering the $\hat{x}$ and $\hat{p}$ quadratures of the bosonic operators $\hat{a}_j,\hat{a}_j^\dagger$. They are Hermitian operators defined by $\hat{x}_j = ({\hat{a}_j + \hat{a}_j^\dagger})/{\sqrt{2}}$ and $\hat{p}_j = {(\hat{a}_j - \hat{a}_j^\dagger)}/{i\sqrt{2}}$ that obey $[\hat{x}_i,\hat{p}_j] = i\delta_{i,j}$. In this basis, \cref{eq:BKC} becomes
\begin{align}
    \hat{\mathcal{H}} &= \sum_j \Big((t+\Delta)\hat{x}_j\hat{p}_{j+1} - (t-\Delta)\hat{x}_{j+1}\hat{p}_j\Big).
\end{align}
As we are interested in studying dynamics, our analysis is centered on the Heisenberg equations of motion
\begin{align}
    \dot{\hat{x}}_j &=  (t+\Delta) \hat{x}_{j-1} - (t-\Delta)\hat{x}_{j+1}, \\
    \dot{\hat{p}}_j &=  (t-\Delta)\hat{p}_{j-1} - (t+\Delta)\hat{p}_{j+1}.
\end{align}
A crucial feature of the system is that the quadrature dynamics are completely decoupled and directional: the $\hat{x}$ chain favors movement to the right and the $\hat{p}$ chain favors movement to the left, without ever mixing. We note that our choice of hopping and squeezing phases in \cref{eq:BKC}, or one that is related by a gauge transformation, is essential to obtain decoupled quadrature dynamics.

As we aim to have the BKC act as a quantum amplifier, it is necessary to study how it behaves when subject to losses, as the latter exist in nearly all platforms that can realize such bosonic quadratic models. In the ideal scenario, only losses associated with the input and output ports are necessary due to the requirement of input-output couplings. However, in realistic physical platforms such as optical systems, each site may be lossy. We treat the loss on each site $j$ as due to Markovian baths through coupling constants $\gamma_j\geq 0$, which is especially relevant to optical systems. These $\gamma_j$ thus correspond to loss rates. Using input-output theory \cite{Clerk2010}, we obtain the Heisenberg-Langevin equations
\begin{align}
    \dot{\hat{x}}_j &= (t+\Delta)\hat{x}_{j-1} - (t-\Delta)\hat{x}_{j+1} -\frac{\gamma_j}{2}\hat{x}_j - \sqrt{\gamma_j}\hat{x}_j^\text{(in)} \label{eq:x_eom_loss}   \\
    &\equiv -i\sum_{\ell=1}^N\big(M_x\big)_{j,\ell}\, \hat{x}_\ell - \sqrt{\gamma_j}\hat{x}_j^\text{(in)} 
\end{align}
and 
\begin{align}
    \dot{\hat{p}}_j &= (t-\Delta)\hat{p}_{j-1} - (t+\Delta)\hat{p}_{j+1} - \frac{\gamma_j}{2}\hat{p}_j - \sqrt{\gamma_j}\hat{p}_j^\text{(in)} \label{eq:p_eom_loss} \\
    &\equiv -i\sum_{\ell=1}^N \big(M_p\big)_{j,\ell}\,\hat{p}_\ell - \sqrt{\gamma_j}\hat{p}_j^\text{(in)},
\end{align}
where $\hat{x}_j^\text{(in)}$ and $\hat{p}_j^\text{(in)}$ are the quadratures of the bath input fields on site $j$ and are the operator equivalent of Gaussian white noise. We note that while there is one physical chain, introducing losses does not couple the $\hat{x}$ and $\hat{p}$ dynamics and we can think of $\hat{x}_j$ and $\hat{p}_j$ as two separate lattices with non-Hermitian dynamics. In the quadrature basis, the full dynamical matrix $M$ is hence block diagonal:
\begin{align}
    M &= \begin{pmatrix}
        M_x & 0 \\
        0 & M_p
    \end{pmatrix},
\end{align}
with $M_x$ and $M_p$ defined above in Eqs.~(\ref{eq:x_eom_loss},\ref{eq:p_eom_loss}). The frequency space quadrature-quadrature susceptibilities in the open-boundary system are defined by
\begin{align}
    \chi^{xx}[\omega] &\equiv -i(\omega\mathds{1}-M_{x,\text{obc}})^{-1}, \label{eq:chixx} \\ 
    \chi^{pp}[\omega] &\equiv -i(\omega\mathds{1}-M_{p,\text{obc}})^{-1}. \label{eq:chipp}
\end{align}
Taking the Fourier transform in time of the Heisenberg-Langevin \cref{eq:x_eom_loss,eq:p_eom_loss} and using the input-output relations $\hat{x}^{(\text{out})}_j = \hat{x}^{(\text{in})}_j + \sqrt{\gamma_j}\hat{x}_j$ and $\hat{p}^{(\text{out})}_j = \hat{p}^{(\text{in})}_j + \sqrt{\gamma_j}\hat{p}_j$, we find
\begin{align}
    \braket{\hat{x}_j^\text{(out)}} &= \braket{\hat{x}_j^\text{(in)}} + \sum_{\ell=1}^N \sqrt{\gamma_j\gamma_\ell} \chi^{xx}[j,\ell;\omega] \braket{\hat{x}_\ell^\text{(in)}} \label{eq:inputoutput_x} \\
    \braket{\hat{p}_j^\text{(out)}} &= \braket{\hat{p}_j^\text{(in)}} + \sum_{\ell=1}^N \sqrt{\gamma_j\gamma_\ell} \chi^{pp}[j,\ell;\omega] \braket{\hat{p}_\ell^\text{(in)}} \label{eq:intputoutput_p}
\end{align}
where the expectation values are taken with respect to a state obeying the Hamiltonian dynamics. The quadrature susceptibilities relate the quadratures of the output fields to those of the input fields and are hence the relevant quantities to study in order to assess the BKC's ability to act as a quantum amplifier. 

We look at how a classical signal $\beta\in\mathds{C}$ sent into the system via the first site is amplified as it propagates towards the last site $N$. To this end, we assume the signal's phase is zero, which amounts to driving the $\hat{x}_1$ quadrature. This is the same setting as in Ref.~\cite{McDonald2020}. We study how the responses on site $j$, given by $\chi^{xx}[j,1;\omega]$ and $\chi^{pp}[j,1;\omega]$, scale with $j$ in the long time limit (or steady state) $\omega=0$. We note that with the diagonal matrix $T$ with $T_{j,j} = (-1)^j$, the dynamical matrix blocks satisfy 
\begin{align}
    \label{eq:trs_obc}
    M_{p,\text{obc}} = TM_{x,\text{obc}}^T T.
\end{align}
The quadrature susceptibilities then obey 
\begin{align}
    \label{eq:quadrature_chi_equivalence}
    \chi^{pp}[j,\ell;\omega] &= (-1)^{j+\ell}\chi^{xx}[\ell,j;\omega].
\end{align}
In other words, as long as no perturbation mixes the $\hat{x}$ and $\hat{p}$ dynamics (such as detunings $\epsilon_j\hat{a}_j^\dagger\hat{a_j}$), the amplification properties of the $\hat{x}$ chain will be the same as that of the $\hat{p}$ chain but in the opposite direction. It thus suffices to look at $\chi^{xx}$. What's more, we assume that the amplitude of the classical drive $\beta$ incident upon the chain is much larger than the average particle number coming from the baths. For our purposes, this is equivalent to having a zero temperature environment, such that no bath particles come into the system. In this setting, the steady-state average particle number on site $j$ satisfies $\bar{n}_j \equiv \braket{\hat{a}_j^\dagger \hat{a}_j}_\text{ss} \simeq \gamma_1\beta^2 |\chi^{xx}[j,1;0]|^2$, where quantum fluctuations due to amplification are neglected. The response $\chi^{xx}$ is then directly linked to the average particle number. As argued in Refs.~\cite{McDonald2020,Lau2018}, the quality of an amplifier is intrinsic to its design and must be independent of the total particle number $\bar{n}_\text{tot} = \sum_{j=1}^N \bar{n}_j$ in the system. Accordingly, it is sufficient to study the spatial distribution of the average particle number; the BKC will be a good quantum amplifier if $\bar{n}_N/\bar{n}_\text{tot}\sim 1$, that is, if particles sent through the first site tend to pile up on the last site in the steady state. 

Since the dynamical matrices $M_{x/p}$ are non-Hermitian, complex conjugation and transposition are no longer equivalent, and the Atland-Zirnbauer classification \cite{AZ1997} is no longer applicable to extract their topological invariant. Instead, we consider the classification of NH Hamiltonians given in Ref.~\cite{Kawabata2019} and find that the BKC with arbitrary on-site dissipation belongs to the class DIII$^\dagger$ which has a $\mathbb{Z}_2$ topological invariant \footnote{This differs from the class AII$^\dagger$ identified in \cite{Busnaina2024} since we choose the hopping and parametric coupling to be imaginary, allowing for the additional particle-hole type symmetry PHS$^\dagger$ \cite{Kawabata2019} given by $-\hat{\mathcal{H}}^* = \hat{\mathcal{H}}$. Nonetheless, both classes have the same $\mathbb{Z}_2$ topological invariant in 1D, see \cite{Busnaina2024}.}. This invariant corresponds to the winding number $\nu$ of the PBC spectral curve $E_{x/p}(k)$ of the dynamical matrix block $M_{x/p}$ about a base point $E_b\in\mathbb{C}$:
\begin{align}
    \label{eq:windingnum_def}
    \nu_{E_b}(E_{x/p}) \equiv \frac{1}{2\pi i}\int_{-\pi}^\pi \; dk \frac{E_{x/p}'(k)}{E_{x/p}(k)-E_b},
\end{align}
where $E'_{x/p}$ is the derivative of $E_{x,p}$ with respect to $k$. Therefore, certain one-dimensional Hamiltonians can have an energy band that forms a closed loop enclosing a nonzero area in the complex plane. Such NH Hamiltonians are said to be point-gapped and are known to give rise to a macroscopic number of edge modes in the OBC system \cite{Okuma2020,Zhang2020,Lee2019,Gong2018,Borgnia2020}. The latter phenomenon is called the non-Hermitian skin effect \cite{Yao2018,Martinez2018}. When the winding number of the PBC curve about the origin $\nu(E_{x/p}) \equiv \nu_{E_b=0}(E_{x/p})$ is nonzero, the OBC edge modes conspire to lead to an amplification (as measured by the susceptibility) that is exponentially large in system size and whose direction is given by the sign of $\nu(E_{x/p})$ \cite{Wanjura2020}. Looking at \cref{eq:quadrature_chi_equivalence} we expect $\nu(E_x) = -\nu(E_p)$. This is confirmed by the fact that \cref{eq:trs_obc} holds under PBC when the total length of the chain $N$ is even. The two PBC spectral curves $E_x$ and $E_p$ are then degenerate and wind in opposite directions. Henceforth, we thus solely concern ourselves with the $\hat{x}$ chain. We note that the dynamical matrix blocks $M_x$ and $M_p$ are unitarily-equivalent to Hatano-Nelson chains \cite{Hatano1996,Hatano1997} \mbox{with on-site loss.}

The connection between the winding number of the PBC spectral curve about the origin and the amplification properties of the system under OBC has been identified as a non-Hermitian bulk-boundary correspondence \cite{Brunelli2023} using the singular value decomposition (SVD) framework outlined in Refs. \cite{Porras2019,Herviou2019}. The SVD of the dynamical matrix $M_x$ is defined as 
\begin{align}
    M_x &= U\Sigma V^\dagger = \sum_j \sigma_j \ket{u_j}\bra{v_j},
\end{align}
where $\Sigma$ is a diagonal matrix and the singular values $\sigma_j =\Sigma_{jj}$ are non-negative and uniquely determined by $M_x$, while $U$ and $V$ are unitary matrices whose columns correspond, respectively, to the left and right singular vectors of $M_x$. The bulk-boundary correspondence is then stated as follows \cite{Brunelli2023}: a nonzero $\nu$ corresponds to $|\nu|$ zero singular values, \textit{i.e.}, singular values $\sigma_j$ of the OBC system that are exponentially small in system size, and whose associated left and right singular vectors are localized (in system size) at opposite edges. These zero singular modes (ZSMs) thus lead to $|\nu|$ channels for directional amplification, as can most transparently be seen from the (formal) steady-state response  
\begin{align}
    \chi^{xx}[\omega=0] &= iM_x^{-1} = \sum_j \frac{i}{\sigma_j}\ket{u_j}\bra{v_j}.
\end{align}
The steady-state response $\chi^{xx}[j,1;0]$ (and hence the steady-state average particle number $\bar{n}_j$) on site $j$ induced by driving the first site then scales exponentially with $j$, provided the amplification channel connects sites 1 and $j$. Such exponential scaling in the susceptibility is dubbed topological amplification \cite{Porras2019}. Naturally, one anticipates this amplification to be robust to some amount of dissipation. We can draw an analogy to disorder in Hermitian systems, where it is expected that the system stays topological as long as the maximal disorder potential is less than the gap size. In the context of lossy, non-Hermitian systems, it was found that the system remains in its topological phase as long as the maximum on-site loss is smaller than the NH gap of the clean system \cite{Wanjura2022}. The NH gap is defined as the minimum distance from the PBC energies to the origin of the complex plane \cite{Brunelli2023}.  In the BKC, we find that in some cases the system can display robustness to loss far beyond this condition.

In this work we study the topological amplification of the BKC subject to dissipation. In particular, we allow for non-uniform, on-site dissipation constants $\gamma_j$ on each site $j$. While the general loss configuration can be studied numerically, we progress by reducing the translation invariance of the lattice by introducing repeated unit cells of length $L$. In this way, the BKC of length $N$ is made up of $N/L$ unit cells whose dissipation constants in Eqs.(~\ref{eq:x_eom_loss},\ref{eq:p_eom_loss}) are $\gamma_1,\gamma_2,\dots,\gamma_L\geq 0$. In the dynamical matrix $M_x$, the loss terms are on the diagonal and satisfy
\begin{align}
    (M_x)_{nL+\ell} = \frac{-i\gamma_\ell}{2}, \quad \ell=1,2,\dots,L
\end{align}
for all $n=0,1,\dots,N/L-1$. See \cref{fig:unit_cell} (a) for a schematic representation. Our numerical and analytical results show that topological amplification can persist even when losses exceed the size of the point gap.

\section{Uniform dissipation, $L=1$}
\label{sec:uniform}
We first study the topological amplification outlined in the previous section in the BKC whose sites are all subject to the same loss, \textit{i.e.}, $\gamma_j = \gamma\geq 0$ for all $j$. While such a situation has been touched upon from a topological standpoint in \cite{McDonald2018,Flynn2020,Wanjura2020,Slim2024,Busnaina2024}, the next sections will generalize the treatment to other dissipative configurations. We thus provide details in the case of uniform dissipation for completeness. 
\begin{figure}[t]
     \centering
     \begin{tabular}{c}
        \begin{subfigure}[t]{0.3\textwidth}
            \centering
            \includegraphics[width=\textwidth]{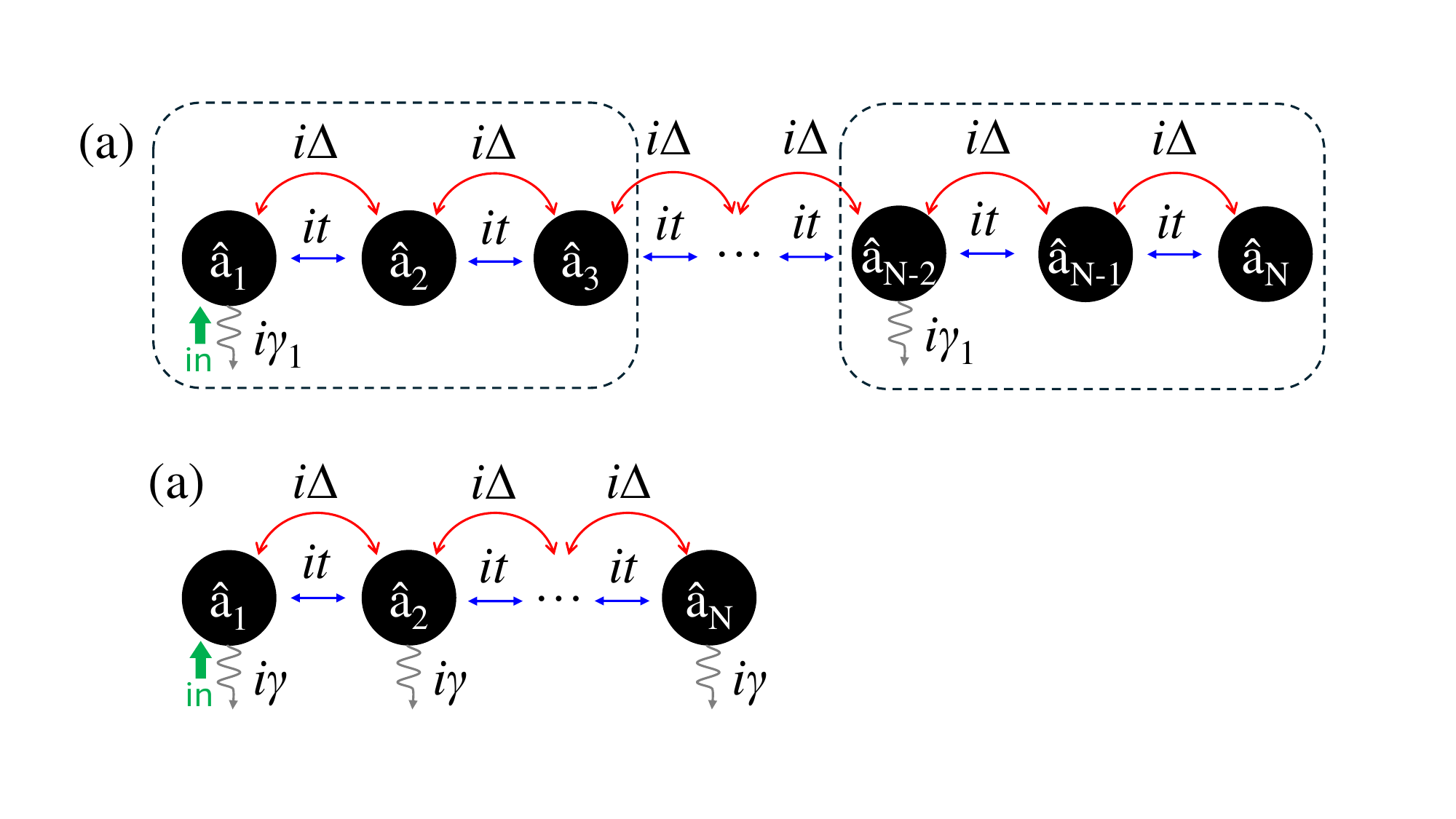}  
        \end{subfigure}
        \\
        \begin{tabular}[b]{ccc}
            \begin{subfigure}{0.158\textwidth}
            \centering
            \includegraphics[width=\textwidth]{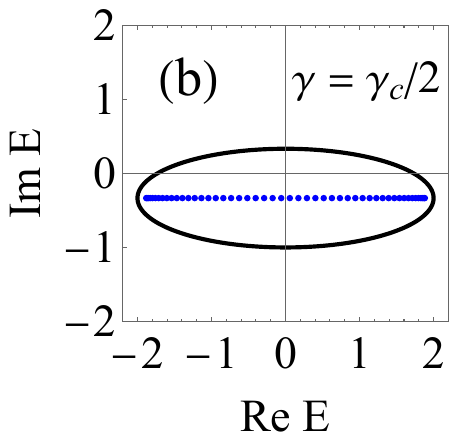}
            \end{subfigure}
            &
            \begin{subfigure}{0.1355\textwidth}
                \centering
                \includegraphics[width=\textwidth]{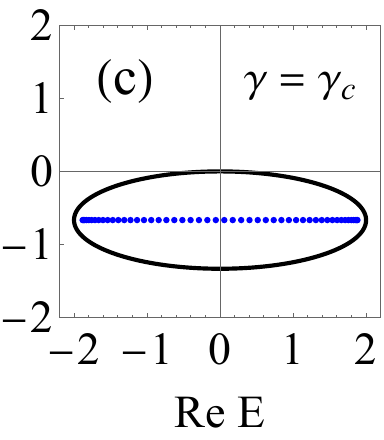}
            \end{subfigure} 
            &
            \begin{subfigure}{0.1355\textwidth}
                \centering
                \includegraphics[width=\textwidth]{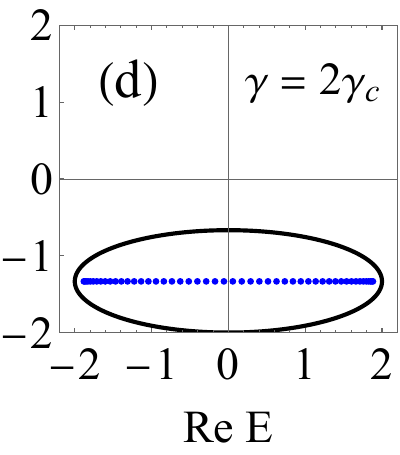}
            \end{subfigure}
        \end{tabular} 
        \\
        \begin{tabular}[t]{cc}
        \begin{subfigure}[t]{0.204\textwidth}
            \centering
            \includegraphics[width=\textwidth]{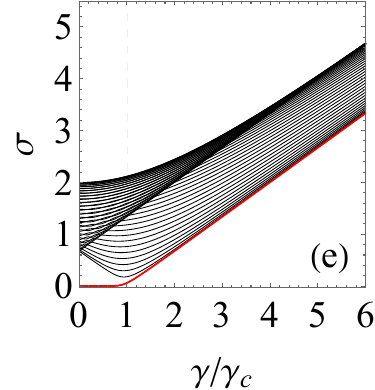}
        \end{subfigure}
        &
        \begin{subfigure}[b]{0.24\textwidth}
            \centering
            \includegraphics[width=\textwidth]{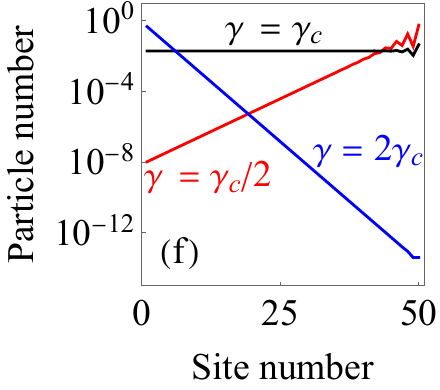}
        \end{subfigure}
        \end{tabular}
     \end{tabular}
    \caption{\justifying(a) Schematic, (b)--(d) spectra, (e) singular values and (f) normalized steady-state particle number distribution for the BKC with 50 sites subject to uniform loss and classical drive on the first site. In \mbox{(b)--(d)}, the PBC (black) and OBC (blue) spectra of $M_x$ are shown for various values of $\gamma$. (e) Singular values of the OBC dynamical matrix $M_x$ for 50 sites. The dashed line corresponds to the critical point $\gamma=\gamma_c$. The smallest singular value is highlighted in red and is responsible for the channel of topological amplification as long as $\gamma < \gamma_c$, after which it merges with the rest of the singular spectrum. In all plots $t=1$ and $\Delta=1/3$.}
    \label{fig:uniform}
\end{figure}

The Hamiltonian $\hat{\mathcal{H}}$ with PBC is translation-invariant and we can therefore write it in momentum space:
\begin{align}
    \label{eq:BKC_uniform}
    \hat{\mathcal{H}} &= \frac{1}{2}\sum_k \begin{pmatrix}
        \hat{a}_k^\dagger & \hat{a}_{-k}
    \end{pmatrix}h(k) \begin{pmatrix}
        \hat{a}_k \\ \hat{a}_{-k}^\dagger
    \end{pmatrix},
\end{align}
where $h(k)$ is the Bloch Hamiltonian and the dynamical matrix $M(k) = \sigma_z h(k) - i\gamma/2$ is
\begin{align}
    \label{eq:dynamical_uniform}
    M(k) &= \left(2t\sin(k) - i\gamma/2\right)\mathds{1} + 2i\Delta \cos(k) \sigma_x
\end{align}
with eigenvalues corresponding to the dynamical matrix blocks $M_{x/p}(k) = 2t\sin(k) + i\left(\pm 2\Delta \cos(k) - \frac{\gamma}{2}\right)$. In this case, increasing $\gamma$ shifts the energies down the imaginary axis while leaving the eigenstates invariant. The PBC spectrum of the dynamical matrix encloses the origin of the complex plane as long as $\gamma < \gamma_c \equiv 4\Delta$. In this case, $\gamma_c/2$ is the point gap of the lossless chain. As $k$ spans the Brillouin zone, $M_x$ winds clockwise around the origin, while $M_p$ winds counterclockwise, that is, $\nu(M_{x/p}) = \mp1$. Each non-trivial winding corresponds to a zero singular value of the OBC system. In turn, they lead to two channels that have opposite directional amplification: one for the $\hat{x}$ chain and the other for the $\hat{p}$ chain. In this setting, a classical signal sent through the first site results in the steady-state average particle number $\bar{n}_j = \braket{\hat{a}_j^\dagger \hat{a}_j}$ being exponentially localized on the last site of the chain. When the dissipation constant is exactly tuned to the critical value $\gamma/2 = 2\Delta$, the PBC spectrum of $M_x$ no longer encloses the origin but touches it at $k=0$. This is a topological phase transition which can be seen as a closing of the NH gap. In this case, the average particle number distribution is uniform along the chain. Increasing $\gamma$ further, the eigenvalues of the dynamical matrix no longer enclose nor touch the origin. The BKC then stops acting as an amplifier and the average particle number distribution localizes on the first site. The three regimes are shown in \cref{fig:uniform} and explicit calculations of the average particle distribution in the steady state are given in \cref{apdx:uniform}.  The critical value can be found by looking for a zero energy state of the PBC spectrum of $M_x$ or by looking for a zero singular value of $M_x$ which is the gap closure of a Hermitian system described by $M_x^\dagger M_x$.  We show this in the next section.

\section{PH Symmetry and PBC Spectral Bands for Chains of $L\geq 1$ Unit Cell Size}
\label{sec:band_separation}

We now turn our attention to systems with reduced translation symmetry. The chain is divided into unit cells of size $L$ and the bath coupling constants $\gamma_j$ may vary within the unit cell. This can lead to the system having more than one PBC energy band. However, regardless of the values of $\gamma_j$, the spectrum of the dynamical matrix block $M_x$ under both OBC and PBC is symmetric about the imaginary axis. That is, $M_x$ obeys the non-Hermitian particle-hole symmetry known as PHS$^\dagger$ \cite{Kawabata2019}
\begin{align}
    \label{eq:phs}
    M_x &= -M_x^*
\end{align}
which follows readily from the fact that $M_x$ has strictly imaginary entries \footnote{Although this seems to come from our choice of gauge, the PHS$^\dagger$ property of \cref{eq:phs} is gauge independent. One can readily derive $M = -M^*$ for the full dynamical matrix from the dynamical equation in the quadrature basis, as $\hat{x}_j$ and $\hat{p}_j$ are Hermitian operators. However, depending on the gauge $\hat{a}_j\to e^{i\theta_j}\hat{a}_j$, $M$ might not be block diagonal in the usual quadrature basis $\hat{x}_j = \frac{1}{\sqrt{2}}(\hat{a}_j + \hat{a}_j^\dagger)$ and $\hat{p}_j = \frac{1}{i\sqrt{2}}(\hat{a}_j - \hat{a}_j^\dagger)$. Instead, the block diagonal basis is $\hat{x}_j = \frac{e^{i\theta}}{\sqrt{2}}(\hat{a}_j + e^{-2i\theta_j}\hat{a}_j^\dagger)$ and $\hat{p}_j = \frac{e^{i\theta}}{i\sqrt{2}}(\hat{a}_j - e^{-2i\theta_j}\hat{a}_j^\dagger)$.}. When there is no dissipation, the spectrum of $M_x$ under OBC is real---which is ensured by the stability assumption, $t>\Delta>0$ \cite{McDonald2018}. The particle-hole symmetry of \cref{eq:phs} hence guarantees a zero-energy mode in the clean system when its total length is odd.  We keep this in mind when analyzing the dissipative chain. 

We now turn our attention to the dissipation in a system made up of $L/N$ unit cells of length $L$ where the dissipation is only on the first site of each unit cell, meaning that $\gamma_1>0$ and $\gamma_j =0$ for $L\geq j \geq 2$. 

\begin{figure}
    \centering
    \begin{tabular}[b]{cccc}
        \begin{subfigure}{0.13\textwidth}
        \centering
        \includegraphics[width=\textwidth]{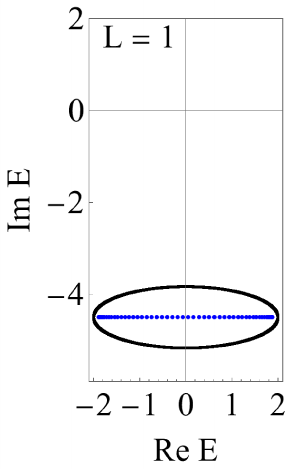}
        \end{subfigure}
        &
        \hspace{-0.25cm}
        \begin{subfigure}{0.11\textwidth}
            \centering
            \includegraphics[width=\textwidth]{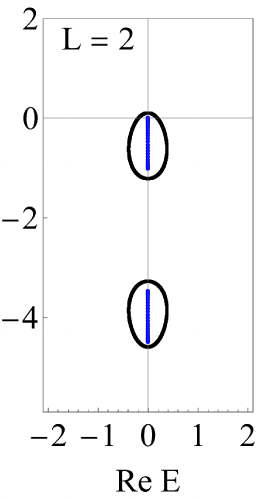}
        \end{subfigure}
        &
        \hspace{-0.25cm}
        \begin{subfigure}{0.11\textwidth}
            \centering
            \includegraphics[width=\textwidth]{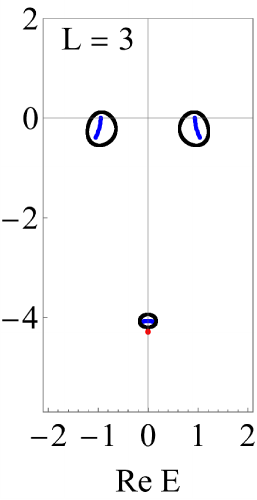}
        \end{subfigure}
        &
        \hspace{-0.25cm}
        \begin{subfigure}{0.11\textwidth}
            \centering
            \includegraphics[width=\textwidth]{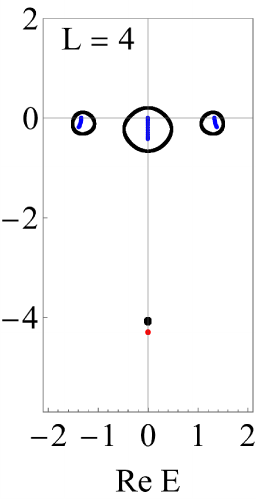}
        \end{subfigure}
    \end{tabular} 
    \caption{\justifying Spectrum of the dynamical matrix $M_x$ for OBC (blue) and PBC (black) for different unit cell lengths $L$ whose first site is subject to loss. When $\gamma_1$ is large enough, the PBC curve splits into $L$ bands. For $L=3,4$, the red point corresponds to the OBC eigenenergy that lies outside of the PBC spectral curve. In all plots $t=1$, $\Delta=1/3$ and $\gamma_1=9$.}
    \label{fig:bandsplitting}
\end{figure}

On the one hand, when $\gamma_1=0$, there is no dissipation and the spectral curve of $M_{x,\text{pbc}}$ winds around the origin. On the other hand, in the asymptotic limit $\gamma_1\to\infty$, the PBC and OBC systems effectively break into $N/L$ copies of a dissipationless BKC with $L-1$ sites and open boundary conditions. One can view this as disconnecting the first site of each unit cell, on which there will be a localized mode of energy $-i\gamma_1/2$. Indeed, as soon as a particle hops on the first site, it directly leaks into the environment, such that neighboring unit cells no longer communicate, see \cref{fig:unit_cell} (b). Since there is no dissipation on the remaining sites, the spectrum of the dynamical matrix block $M_{x,\text{obc}}$ associated with the BKC of length $L-1$ is real. When $L-1$ is odd, the PHS$^\dagger$ of \cref{eq:phs} ensures the existence of a zero energy mode. Therefore, as $\gamma_1$ is increased from zero, $N/L$ eigenenergies of $M_x$ tend to the origin of the complex plane. Since the spectrum of $M_{x,\text{pbc}}$ winds around the origin at $\gamma_1=0$, we anticipate the PBC spectral curve to keep winding around the origin as $\gamma_1$ is increased when $L$ is even, but not when $L$ is odd. Additionally, $N/L$ energies tend to $-i\infty$ as $\gamma_1\to\infty$. Using Gershgorin's circle theorem \cite{Gershgorin1931} on $M_{x,\text{pbc}}$, we find that the PBC spectral curve corresponds to at least two bands when $\gamma_1>8t$. In fact, we numerically observe that the PBC spectral curve of the BKC with unit cell length $L$ splits into $L$ bands when $\gamma_1$ is large enough, see \cref{fig:bandsplitting}. 

No matter the number of bands or whether the PBC spectral curve winds around the origin, we numerically observe the NH skin effect when $\gamma_1$ is small enough. That is, the spectrum of $M_{x,\text{obc}}$ lies inside the PBC spectral curve and all of the OBC eigenstates are localized on the last site of the chain. However, for $L>2$, as $\gamma_1$ is increased past a critical value $\zeta>0$, there is a single OBC eigenenergy that leaves the area enclosed by the PBC spectral curve, see the red points in \cref{fig:bandsplitting}. In more rigorous terms, the winding number of the PBC spectral curve about this OBC eigenenergy changes when $\gamma_1=\zeta$. Furthermore, the associated OBC eigenstate transitions from a NH skin mode localized on the last site when $\gamma_1<\zeta$, to a completely delocalized state at $\gamma_1=\zeta$ and to an edge state localized on the first site when $\gamma_1>\zeta$. The details of this OBC mode are provided in \cref{apdx:edge_transition}. We note that the behavior of this mode is not tied to the topological amplification.

\section{Topological Amplification, Topological Invariants and the BKC with Reduced Periodicity}\label{sec:robutness}
\begin{figure}[t]
     \centering
    \begin{tabular}[b]{c}
        \hspace{1.2cm}
        \begin{subfigure}{0.4\textwidth}
            \centering
            \includegraphics[width=\textwidth]{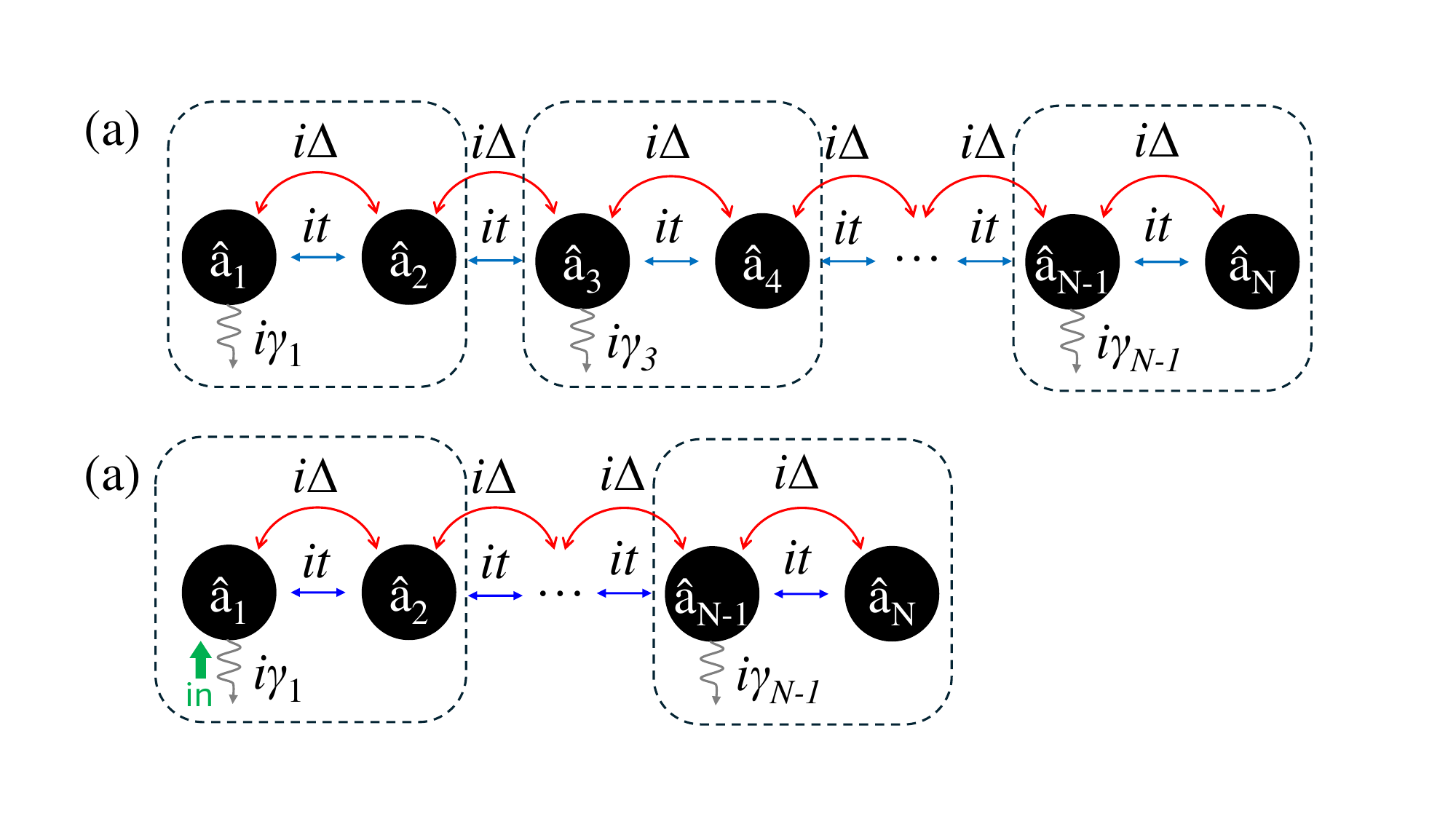}
        \end{subfigure}
        \\
        \hspace{-0.4cm}
        \begin{tabular}{cc}
             \begin{subfigure}{0.265\textwidth}
            \centering
            \includegraphics[width=\textwidth]{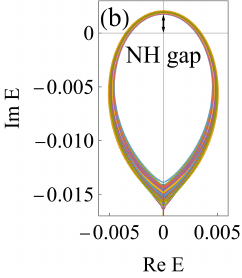}
        \end{subfigure}
        &
        \hspace{-0.4cm}
        \begin{subfigure}{0.223\textwidth}
            \centering
            \includegraphics[width=\textwidth]{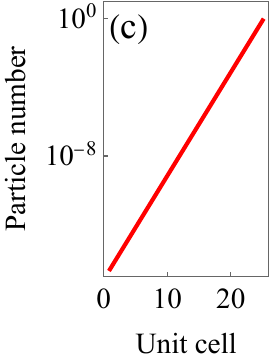}
        \end{subfigure}
        \end{tabular}
    \end{tabular}
    \caption{ \justifying (a) Schematic of the BKC with loss on its odd sites. (b) Spectrum of the dynamical matrix under PBC for the BKC with 1000 sites (500 unit cells) whose odd sites exhibit loss sampled from the uniform distribution $\gamma_{2j+1}/t\in[0,1000]$. Each color of points represents one of the 100 runs with a different realization of disorder. The system is always in a nontrivial topological phase; the NH gap remains open for all 100 runs with the origin inside the energy band. (c) Normalized steady-state particle number distribution $\braket{\hat{a}_j^\dagger \hat{a}_j} \propto |\chi^{xx}[j,1;\omega=0]|^2$ for the BKC under open boundary conditions (OBC) with 50 sites (25 unit cells). The particle distribution is exponentially localized on the last unit cell and is independent of $\gamma_1,\gamma_3,\dots,\gamma_{N-1}$.}
    \label{fig:L2}
\end{figure}

\begin{figure*}[ht!]
     \centering
    \begin{tabular}[b]{ccccc}
        \begin{subfigure}{0.2035\textwidth}
        \centering
        \includegraphics[width=\textwidth]{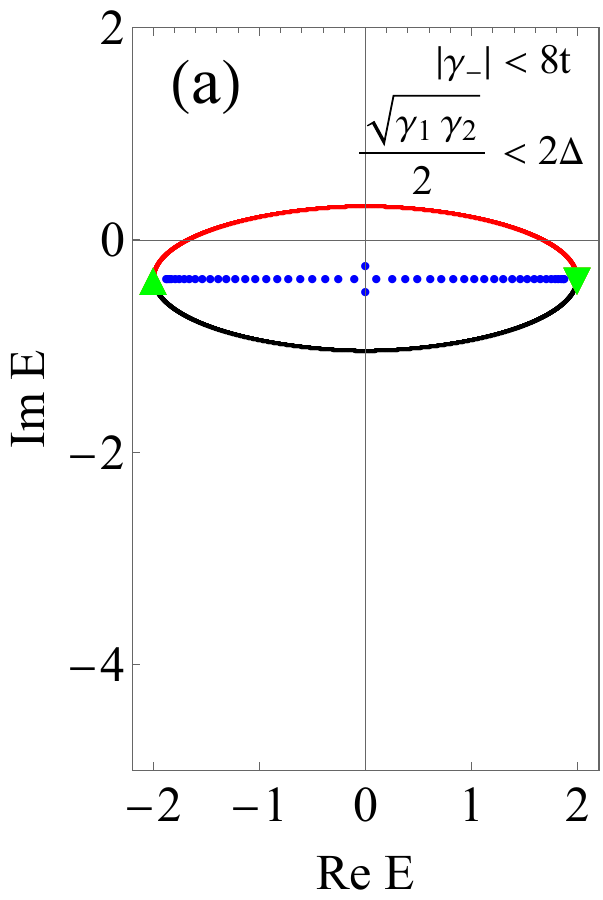}
        \end{subfigure}
        &
        \begin{subfigure}{0.18\textwidth}
            \centering
            \includegraphics[width=\textwidth]{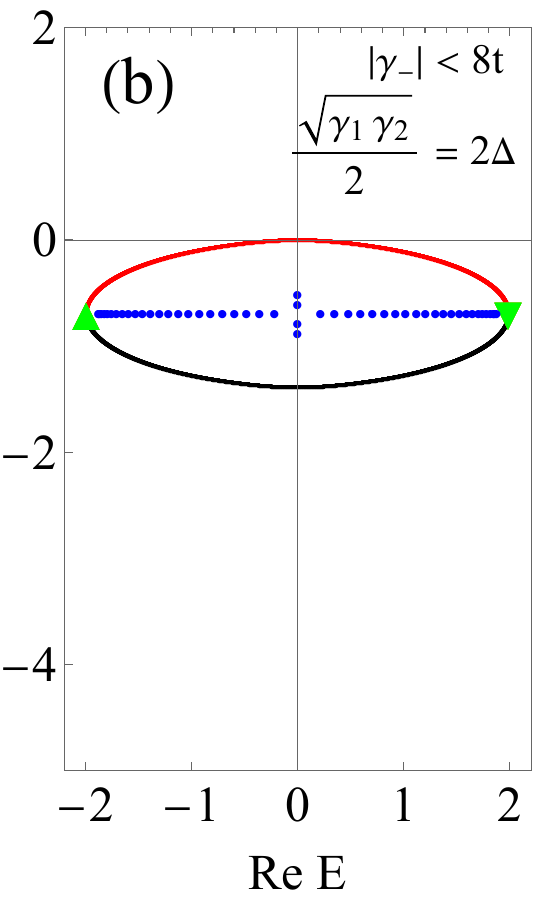}
        \end{subfigure} 
        &
        \begin{subfigure}{0.18\textwidth}
            \centering
            \includegraphics[width=\textwidth]{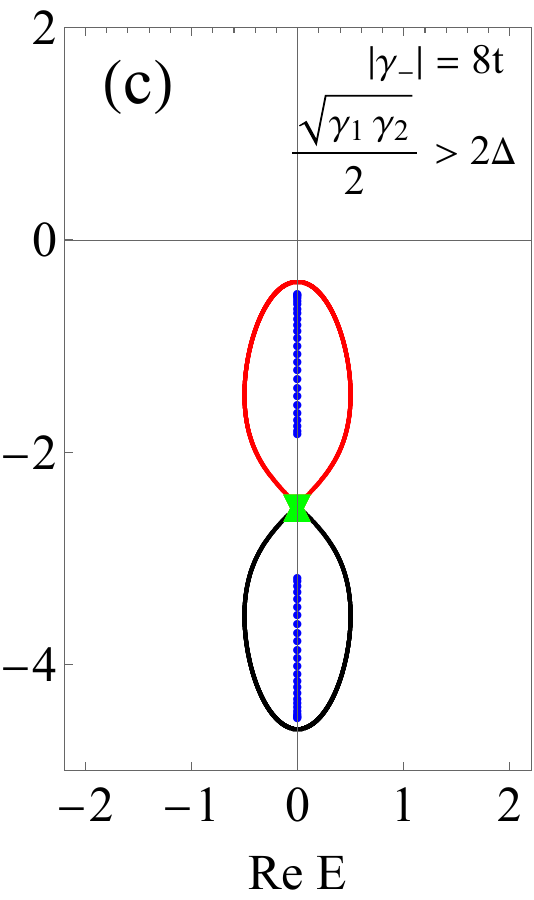}
        \end{subfigure}
        &
        \begin{subfigure}{0.18\textwidth}
            \centering
            \includegraphics[width=\textwidth]{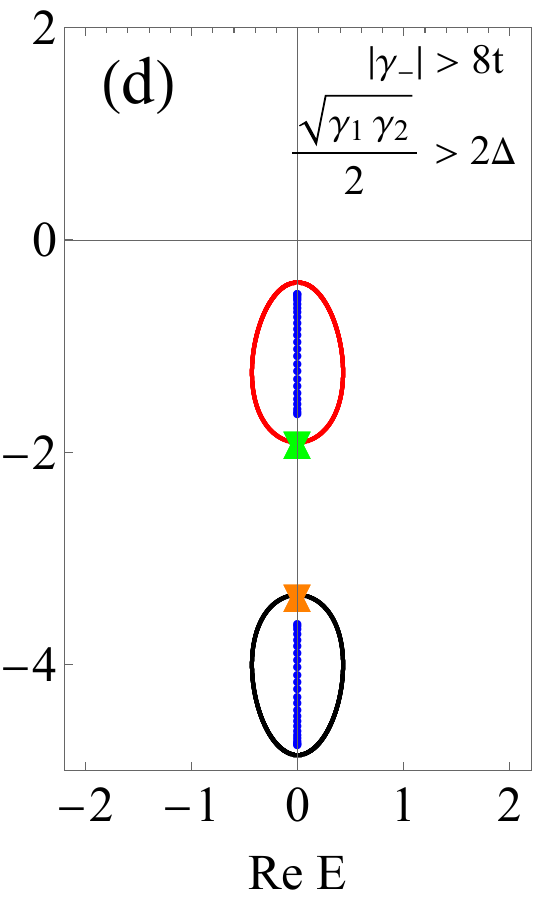}
        \end{subfigure}
        &
        \begin{subfigure}{0.18\textwidth}
            \centering
            \includegraphics[width=\textwidth]{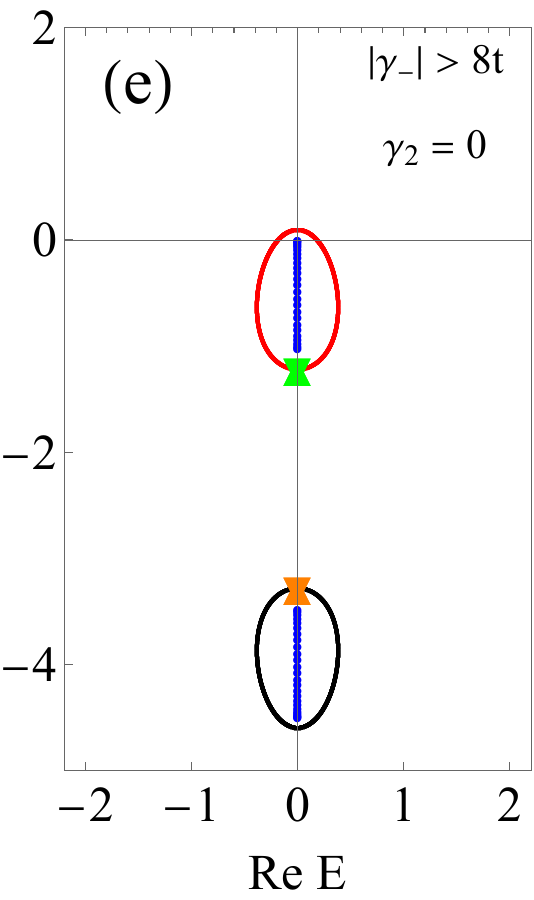}
        \end{subfigure}
    \end{tabular} 
    \caption{ \justifying Spectra of the dynamical matrix block $M_x$ for the two-site unit cell BKC for various values of bath coupling constants $\gamma_1\geq 0$ and $\gamma_2\geq 0$. The OBC spectrum (blue) is for a chain of 50 sites. The red (black) curve corresponds to the part of the PBC spectrum given by $E_+$ ($E_-$), see \cref{eq:L2_Epm}. The PBC spectrum is subject to two independent processes: it winds around the origin depending on how $\frac{\sqrt{\gamma_1\gamma_2}}{2}$ relates to $2\Delta$, while the presence of a line gap is associated with how $|\gamma_-| \equiv |\gamma_1 - \gamma_2|$ compares to $8t$. Further, $\color{green}\blacktriangle$ corresponds to $\lim_{k\downarrow-\pi}E_+(k)$ and $\color{green}\blacktriangledown$ corresponds to $\lim_{k\uparrow\pi}E_+(k)$. When visible, $\color{orange}\blacktriangle$ corresponds to $\lim_{k\uparrow\pi}E_-(k)$ and $\color{orange}\blacktriangledown$ corresponds to $\lim_{k\downarrow-\pi}E_-(k)$. Otherwise, $\color{green}\blacktriangle$ overlaps with $\color{orange}\blacktriangle$ and $\color{green}\blacktriangledown$ overlaps with $\color{orange}\blacktriangledown$. See \cref{subsec:even_unit_cell} for details. Panels (a)--(d) have $\gamma_2=1$ while (e) has $\gamma_2=0$. (a) $\gamma_1/2 = 2\Delta^2$. (b) $\gamma_1/2 = 4\Delta^2$. (c) $\gamma_1 = 9$. (d) $\gamma_1 = 9.5$. (e) $\gamma_1 = 9$. In all plots $t=1$ and $\Delta=1/3$.}
    \label{fig:L2_spectra}
\end{figure*}

\begin{figure}[ht!]
    \begin{tabular}{cc}
    \hspace{-0.5cm}
        \begin{subfigure}[t]{0.26\textwidth}
            \centering
            \includegraphics[width=\textwidth]{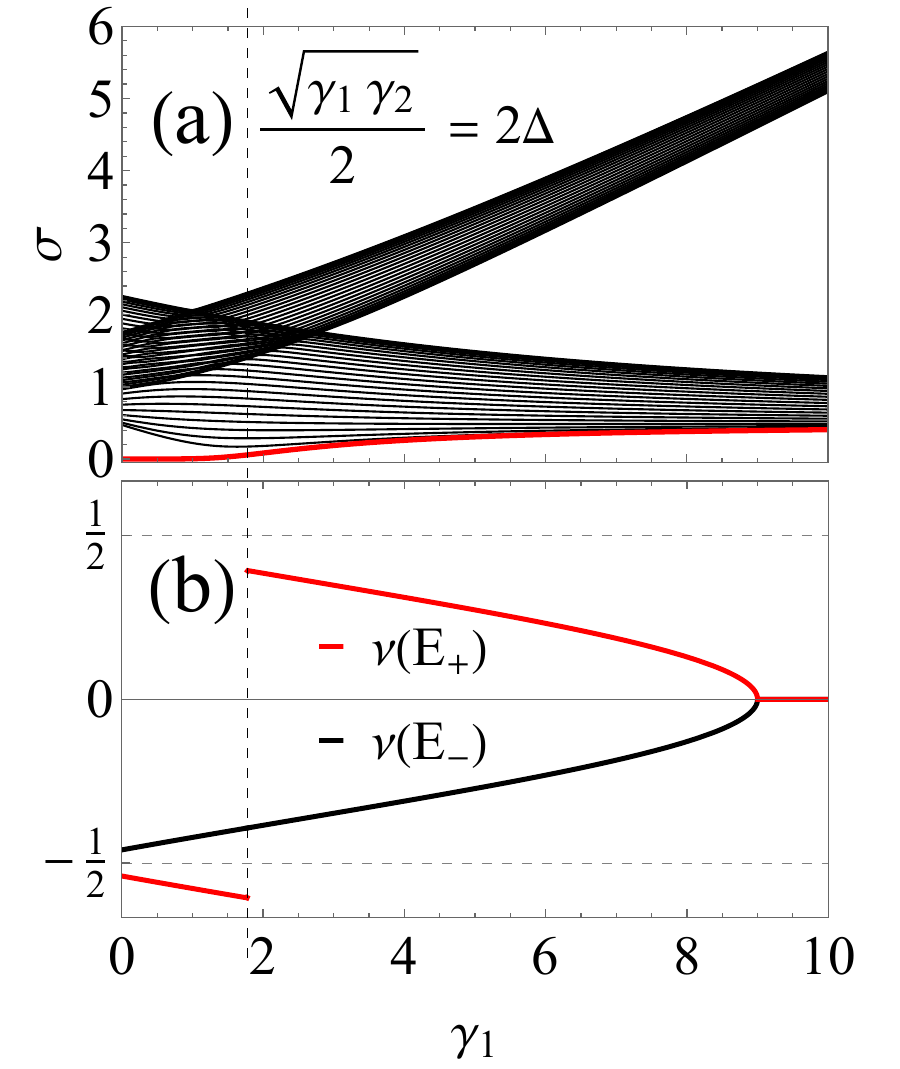}
        \end{subfigure} 
        &
        \hspace{-0.7cm}
        \begin{subfigure}[t]{0.26\textwidth}
            \centering
            \includegraphics[width=\textwidth]{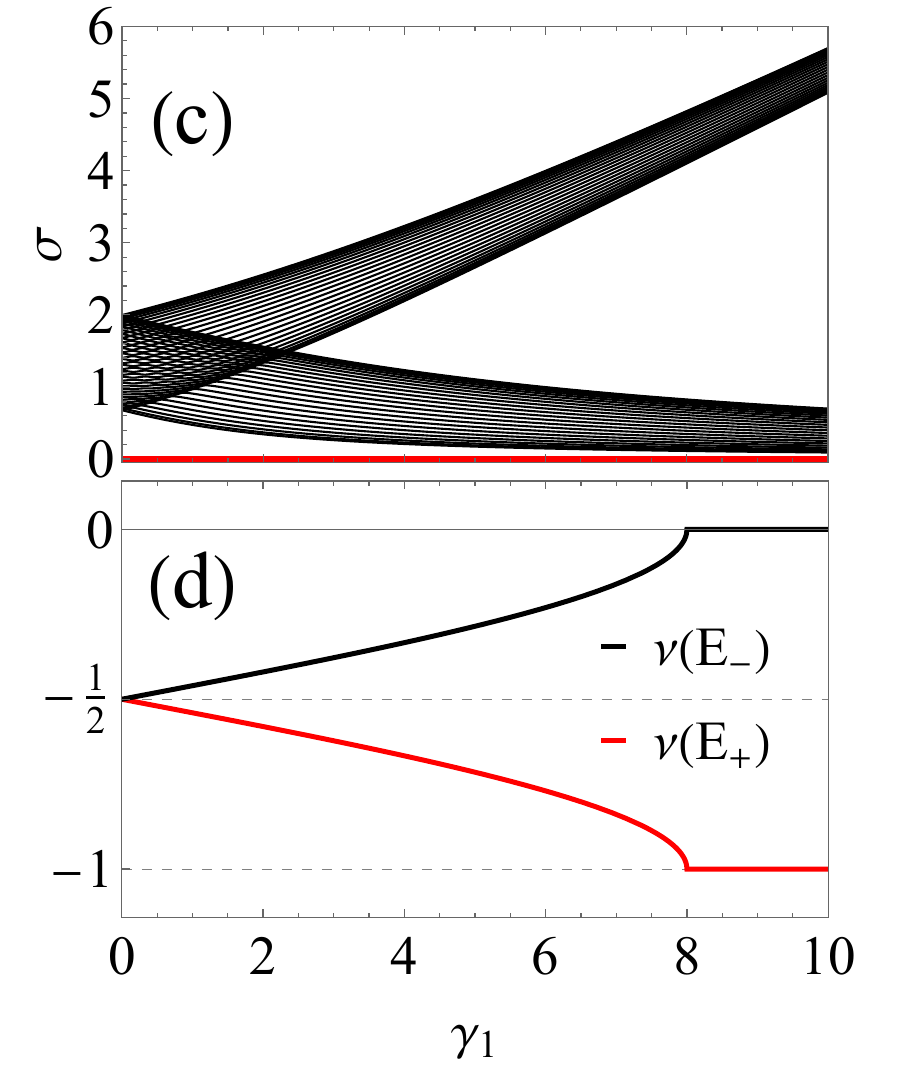}
        \end{subfigure}
    \end{tabular}
    \caption{\justifying Singular values [(a) and (c)] and winding number contributions [(b) and (d)] of the two-site unit cell BKC dynamical matrix block $M_x$ as a function of the bath coupling constant $\gamma_1$, see \cref{fig:L2_spectra} for the spectra. The singular values $\sigma$ are those of the 50-site open BKC, with the smallest singular value in red. (a)--(b) The unit cell's second site has $\gamma_2=1$ allowing for a topological phase transition at $\frac{\sqrt{\gamma_1\gamma_2}}{2} = 2\Delta$, depicted by the vertical dashed line. The emergence of a line gap in the PBC system happens at $\gamma_1=9$, at which point the absolute difference in losses satisfies $|\gamma_-| = 8t$. (c)--(d) The unit cell's second site shows no loss $\gamma_2=0$. In this case, the system exhibits no topological phase transition; the winding number $\nu(E) = \nu(E_+) + \nu(E_-)=-1$ for all $\gamma_1\geq 0$ and the emergence of a line gap happens at $\gamma_1=8$. See \cref{subsec:even_unit_cell} for more details. In all plots $t=1$ and $\Delta=1/3$.}
    \label{fig:L2_singular}
\end{figure}

As a hallmark of non-trivial topological phases, the channels of exponential amplification are generally robust to disorder for driven-dissipative systems. This is guaranteed provided that the size of the non-Hermitian gap is larger than the maximum amount of disorder present in the system \cite{Wanjura2022}. While this is a sufficient condition, we show that in certain cases, the BKC exhibits exponential amplification even if the maximum disorder value exceeds the NH gap. We find that when the total length of the chain $N$ is even, the nontrivial topological phase responsible for exponential amplification remains for arbitrarily large loss on odd sites $\gamma_{2j+1}$, see \cref{fig:L2}. Moreover, we show that the intuition outlined in the previous section is borne out: when $L$ is odd, dissipation on the first site of the unit cells is enough to induce a topological phase transition, while this is not true when $L$ is even.  This is demonstrated by observing the NH spectrum of the system with $\gamma_1 \geq 0$ as seen in Figs.~[\ref{fig:L2_spectra},\ref{fig:L2_singular}] for $L=2$, and in Fig.~\ref{fig:L3} for $L=3$.  Without dissipation the spectrum winds around the origin.  Then upon increasing $\gamma_1$ the spectrum splits into $L$ disconnected curves.  These curves become smaller as $\gamma_1$ is increased and winds around their $\gamma_1\to\infty$ value.  This means one curve winds around $-i\gamma/2$ and $L-1$ curves wind around the (real) solutions of a non-dissipative BKC with $L-1$ sites with open boundary conditions. It is therefore expected that when $L-1$ is odd (even unit cell length) one of the disconnected curves will always wind around the origin as the clean $L-1$ chain with OBC has a zero energy solution.


\subsection{Unit Cell With $L=2$ Sites}\label{subsec:even_unit_cell}

To explain the surprising robustness to disorder depicted in \cref{fig:L2}, we first consider the case where the unit cell of the BKC has two sites. In momentum space, the block matrix $M_x(k)$ is
\begin{align}
\begin{split}
    &M_x(k) = \\
    &\begin{pmatrix}
        \frac{-i\gamma_1}{2} & i(t+\Delta)e^{-ik} - i(t-\Delta) \\ 
        i(t+\Delta) - i(t-\Delta) e^{ik} & -\frac{i\gamma_2}{2}
    \end{pmatrix},
    \end{split}
\end{align}
where $k\in(-\pi,\pi)$ is confined to the reduced Brillouin zone and we set the lattice constant $2a=1$. Of course, taking $\gamma_1=\gamma_2$ reduces to the case of uniform dissipation. The eigenvalues are found to be 
\begin{align}
\begin{split}
    E_\pm(k) &= -\frac{i\gamma_+}{4} \label{eq:L2_Epm} \\
    &\hspace{-37pt}\pm i\sqrt{\frac{\gamma_-^2}{16} - 2(t^2-\Delta^2) + (t+\Delta)^2e^{-ik} + (t-\Delta)^2 e^{ik}}.
    \end{split}
\end{align}
where $\gamma_\pm \equiv \gamma_1\pm \gamma_2$. Importantly, the curves $E_+$ and $E_-$ do not constitute parts of a single closed curve for all parameter values. We already know that they form a closed loop when $\gamma_1 = \gamma_2$, see \cref{fig:uniform}. However, the situation becomes very different when $\gamma_-\neq 0$. Analysis of the spectrum in Eq.~\ref{eq:L2_Epm} shows that the solutions split into two curves when $|\gamma_-| > 8t$ as can be seen in Fig~\ref{fig:L2_spectra}.  In this region the spectrum develops a line gap.
We remark that when $\gamma_2=0$, the critical value $\gamma_- = \gamma_1 = 8t$ is exactly that identified using the Gershgorin circle theorem \cite{Gershgorin1931} estimate in \cref{sec:band_separation}. Since a line gap is defined as a line in the complex-energy plane that does not cross the spectrum and has energy values on either side, the critical value $|\gamma_-|=8t$ separates a regime where there is no line gap in the PBC system from one where there is. However, in our setting, the emergence of such a line gap is not topological. Indeed, even when the energy levels $E_\pm$ do not form closed curves on their own, we can still compute their contribution to the winding number $\nu(E_\pm)$ using \cref{eq:windingnum_def}, which obey $\nu(E) = \nu(E_+) + \nu(E_-)$. Using the argument principle, the winding number $\nu(E)$ satisfies $\nu(E) = Z-P$, where $Z$ and $P$ are the number of zeroes and poles of 
\begin{align}
    \begin{split}
    E(z) &= -\frac{i\gamma_+}{4} \\
    &\hspace{-10pt}\pm \frac{i}{4}\sqrt{\gamma_-^2 - 32(t^2-\Delta^2) + \frac{16(t+\Delta)^2}{z} + 16(t-\Delta)^2z}
    \end{split}
\end{align}
within the unit circle, respectively. There is evidently a pole at $z=0$. For the zeroes, we find 
\begin{align}
    z_\pm &= \frac{\gamma_1\gamma_2 + 8(t^2-\Delta^2) \pm \sqrt{\gamma_1^2\gamma_2^2 + 16\gamma_1\gamma_2(t^2-\Delta^2)}}{8(t-\Delta)^2}.
\end{align}
Although we always have $|z_+| > 1$ as long as $0<\Delta<t$ and $\gamma_1\gamma_2>0$, the zero $z_-$ lies inside the unit circle when $\frac{\sqrt{\gamma_1\gamma_2}}{2} > 2\Delta$, and outside when $\frac{\sqrt{\gamma_1\gamma_2}}{2} < 2\Delta$. Thus
\begin{align}
    \label{eq:winding_number_L2}
    \nu(E) &= \begin{dcases}
        -1 & \frac{\sqrt{\gamma_1\gamma_2}}{2} < 2\Delta, \\
        0 & \frac{\sqrt{\gamma_1\gamma_2}}{2} > 2\Delta.
    \end{dcases}
\end{align}
The value $\frac{\sqrt{\gamma_1\gamma_2}}{2}=2\Delta$ is hence identified as the point at which the NH gap closes and can also be found by requiring a zero mode in the PBC spectrum or singular spectrum. When $\gamma_2=0$ ($\gamma_1=0$), the PBC spectral curve always winds around the origin and $\nu(E) = -1$ for all $\gamma_1\geq 0$ ($\gamma_2\geq 0$), as shown in \cref{fig:L2_spectra} (e). 
We conclude that there are two processes at play simultaneously: the emergence of a line gap at $|\gamma_-|=8t$, see \cref{fig:L2_spectra} (c)-(d), and the topological phase transition at $\frac{\sqrt{\gamma_1\gamma_2}}{2} = 2\Delta$, see \cref{fig:L2_spectra} (b). Interestingly, these two processes are independent of each other since one depends exclusively on $t$ and the other on $\Delta$. As long as there is a PBC energy band winding around the origin, the system exhibits topological amplification regardless of whether there exists a line gap. 
We note that this is a consequence of our setting and may not always be true; in \cref{apdx:gainloss_balance}, we show that the line gap can be topological if we add gain on every site.
When $\frac{\sqrt{\gamma_1\gamma_2}}{2} < 2\Delta$, the BKC with a two-site unit cell hosts two zero singular modes, one associated with the $\hat{x}$ chain and the other with the $\hat{p}$ chain, see \cref{fig:L2_singular}. In turn, these ZSMs lead to two channels for exponential amplification whose directions are opposite. This means that any $\gamma_1$, as large as it may be, can be compensated by reducing $\gamma_2$ to remain in the topological phase of exponential amplification. For completeness, we analytically verify that there is exponential amplification when $\frac{\sqrt{\gamma_1\gamma_2}}{2} < 2\Delta$ using the system's susceptibility in \cref{apdx:L2}. At the critical value $\frac{\sqrt{\gamma_1\gamma_2}}{2} = 2\Delta$, the $\hat{x}$ chain hosts a zero energy eigenstate of $M_{x,\text{pbc}}$ with zero momentum. It is readily found to be $\ket{\psi_0} = \bigotimes_{j=1}^{N/2}\ket{\phi_0}$, with the intra-unit cell vector $\ket{\phi_0} = \frac{1}{\mathcal{N}}(\gamma_2/4\Delta, 1)^T$ with normalizing constant $\mathcal{N} = \sqrt{1+\gamma_2^2/\gamma_1^2}$. 

The fact that the system exhibits exponential amplification for all values of $\gamma_1$ as long as $\gamma_2=0$ far exceeds the expectation that a topological phase transition will occur when the maximum disorder value surpasses the size of the NH gap \cite{Wanjura2022}. The topological amplification is then insensitive to the losses on the chain's odd sites. In fact, we may as well allow for arbitrary bath coupling constants $\gamma_1,\gamma_3,\dots,\gamma_{N-1}$ on the odd sites of the chain. This is the essence behind \cref{fig:L2} (b), where the dissipation on the even sites is zero while the dissipation on the odd sites is random and yet, the spectrum winds around the origin. To understand why this generalization works, consider the spectrum of the periodic dynamical matrix blocks $M_x$. When the odd bath coupling constants $\gamma_1,\gamma_3,\dots,\gamma_{N-1}\geq 0$ are small enough, the PBC spectrum winds around the origin and the OBC system exhibits exponential amplification. For a topological phase transition to occur for certain values of odd bath coupling constants, the PBC spectral curve must cross the origin and a zero energy mode must exist. That is, there must be some state $\ket{\psi_0}=(\psi_1,\dots,\psi_N)$ such that $M_x\ket{\psi_0} = 0$. Explicitly, 
\begin{subequations}
\begin{align}
    i(t+\Delta)\psi_{2j} - i(t-\Delta)\psi_{2j+2} - \frac{i\gamma_{2j+1}}{2}\psi_{2j+1} &= 0 \label{eq:Leven_dynamicaleq_1}\\
   i(t+\Delta)\psi_{2j-1} - i(t-\Delta) \psi_{2j+1} &= 0, \label{eq:Leven_dynamicaleq_2}
\end{align}
\end{subequations}
for $1\leq j \leq \frac{N-2}{2}$. However, one can prove that such a state cannot exist. Indeed, letting $r>0$ be defined by $e^{2r} = \frac{t+\Delta}{t-\Delta}$, \cref{eq:Leven_dynamicaleq_2} yields $\psi_{2j+1} = e^{2r}\psi_{2j-1}$. Since $N-1$ is odd, we find $\psi_{N-1} = e^{r(N-2)}\psi_1$. However, the PBC gives $\psi_1 = e^{2r-ik}\psi_{N-1}$ meaning $\psi_{2j+1} = 0$ for all $j$. Thus, \cref{eq:Leven_dynamicaleq_1} yields $\psi_{2j+2} = e^{2r}\psi_{2j}$ and, similarly, we find $\psi_{2j}=0$. Therefore, the fact that the system's dynamics are non-reciprocal and that the eigenstates are exponentially localized on the edge under OBC is exactly what prevents the PBC spectrum from containing a zero energy eigenstate. We conclude that, no matter the values of the odd losses, $M_{x,\text{pbc}}$ never has a zero energy eigenstate, and hence that the PBC spectral curve never crosses the origin as $\gamma_1,\gamma_3,\dots,\gamma_{N-1}$ are increased from zero. The system is thus always in a nontrivial topological phase. In other words, no matter the on-site losses that are introduced on the odd sites, the system whose length is even will always exhibit exponential amplification. We prove this statement using the susceptibility via a repeated application of Dyson's equation in \cref{apdx:Neven_robustness_proof}.

\subsection{Unit Cell With $L>2$}\label{subsec:general_L}

We now consider the BKC with a unit cell of more than two sites and with dissipation only on its first site. We look at whether there is a specific $\gamma_1>0$ at which dissipation will induce a topological phase transition out of the phase of exponential amplification. To this end, we separate the general situation $L>2$ into two cases: one where $L$ is even and one where $L$ is odd. 

\subsubsection{Even $L$}\label{subsubsec:even_unit_cell}

When the length of the unit cell is even, the total length of the chain $N$ is always even, regardless of the number of unit cells $N/L$. Therefore, a chain with even $L$ and nonzero dissipation $\gamma_1$ on the first site of its unit cells is a chain with an even length and nonzero dissipation on certain odd sites. From the previous section, we know that, as long as $\gamma_2=\gamma_4=\dots=\gamma_N=0$, no amount of loss on the odd sites of a chain with even length can induce a topological phase transition out of the exponentially-amplifying phase. Therefore, for any even $L$, there will be topological amplification for any amount of loss $\gamma_1>0$. This confirms the intuition provided in \cref{sec:band_separation}. 

\subsubsection{Odd $L$}\label{subsubsec:odd_unit_cell}

\begin{figure}[t!]
     \centering
     \begin{tabular}{c}
        \hspace{0.2cm}
        \begin{subfigure}[t]{0.43\textwidth}
            \centering
            \includegraphics[width=\textwidth]{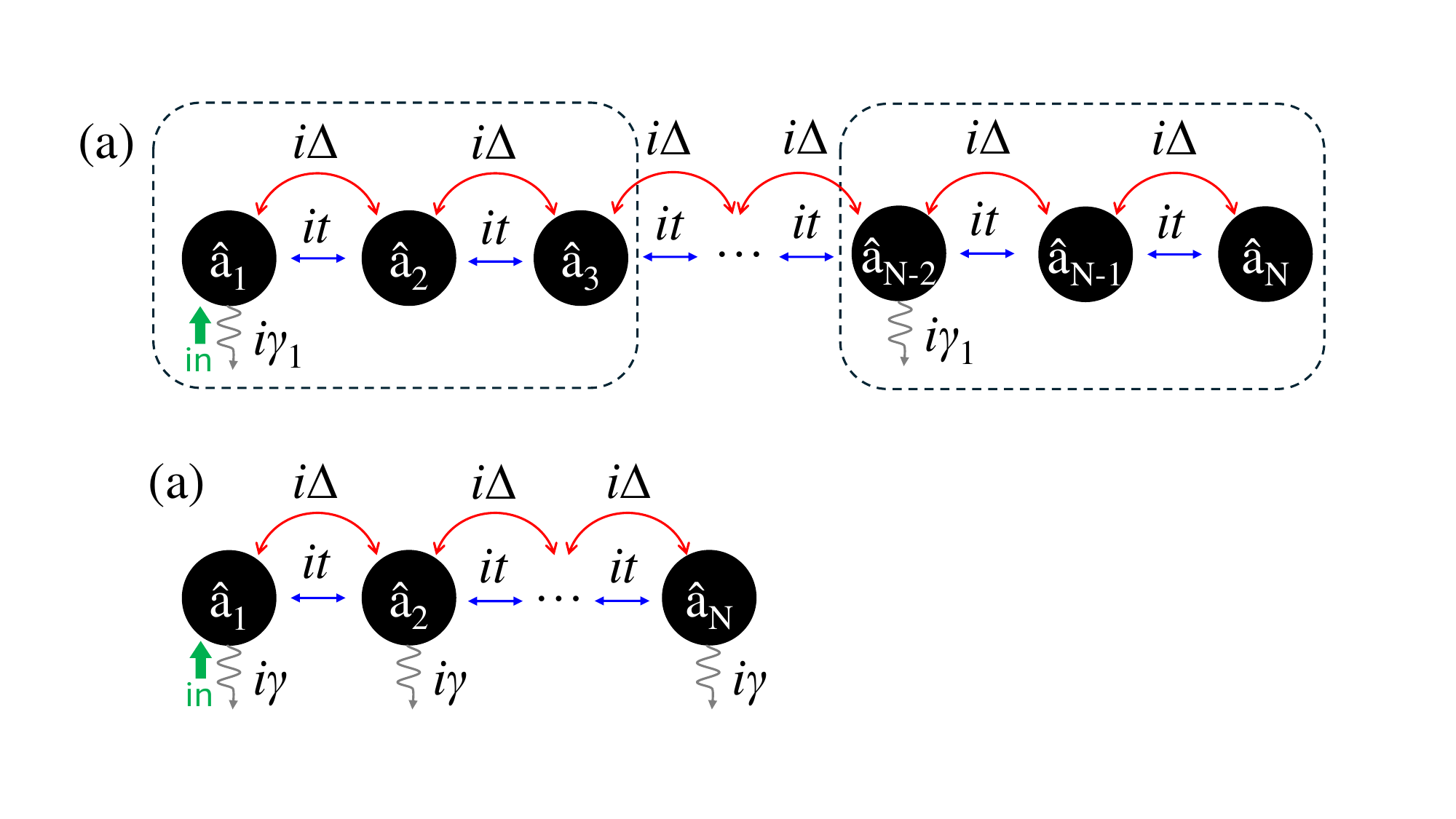}
        \end{subfigure} 
        \\
        \begin{tabular}[b]{ccc}
            \begin{subfigure}{0.1575\textwidth}
            \centering
            \includegraphics[width=\textwidth]{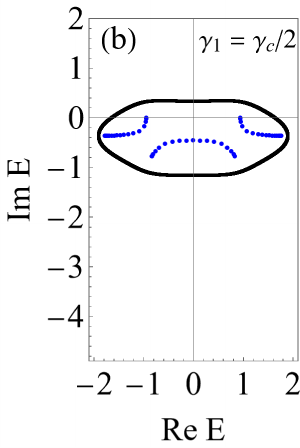}
            \end{subfigure}
            &
            \begin{subfigure}{0.1355\textwidth}
                \centering
                \includegraphics[width=\textwidth]{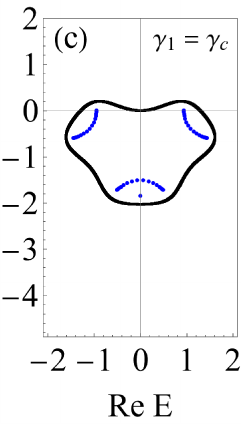}
            \end{subfigure} 
            &
            \begin{subfigure}{0.1355\textwidth}
                \centering
                \includegraphics[width=\textwidth]{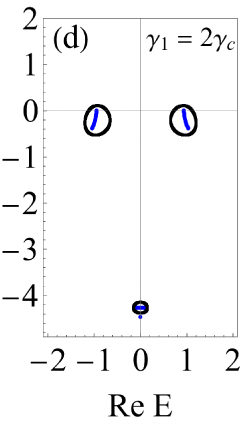}
            \end{subfigure}
        \end{tabular} 
        \\
        \begin{tabular}{cc}
             \begin{subfigure}[t]{0.21\textwidth}
            \centering
            \includegraphics[width=\textwidth]{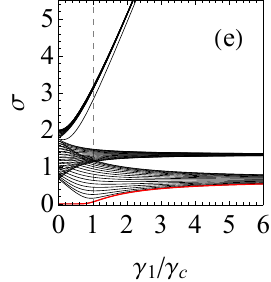}
        \end{subfigure}
        \begin{subfigure}[t]{0.23\textwidth}
            \centering
            \includegraphics[width=\textwidth]{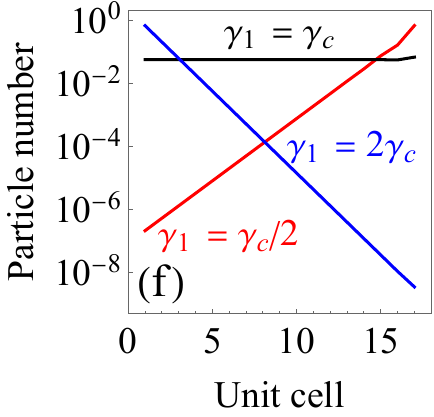}
        \end{subfigure}
        \end{tabular}
     \end{tabular}
    \caption{\justifying (a) Schematic, (b)--(d) spectra, (e) singular values and (f) normalized steady-state unit cell particle number distribution for the three-site unit cell BKC with 51 sites (17 unit cells) and loss on every three sites. In (b)--(d), the PBC (black) and OBC (blue) spectrum of $M_x$ are shown for different values of the bath coupling constant $\gamma_1$, showcasing a topological phase transition. (e) The singular values are plotted for the dynamical matrix block $M_x$ of the (open) BKC with 51 sites as a function of the bath coupling constant $\gamma_1$. The critical point $\gamma_1=\gamma_c$ is depicted by the vertical dashed line and separates a regime wherein the smallest singular value (red) is exponentially small in system size from one in which it is not. See \cref{subsubsec:odd_unit_cell} for details. In all plots $t=1$ and $\Delta=1/3$, so that the critical value \cref{eq:g1_criticalvalue} yields $\gamma_c=14/3$.}
    \label{fig:L3}
\end{figure}

The situation when $L$ is odd is drastically different from when $L$ is even; we find that one lossy site per unit cell is sufficient to induce a topological phase transition, see \cref{fig:L3}. When $\gamma_1$ is small enough, the system is in a non-trivial topological phase, with the spectrum of the dynamical matrix winding around the origin [\cref{fig:L3} (b)]. In this setting, there is a zero singular mode of $M_{x,\text{obc}}$ whose singular value, depicted in red in \cref{fig:L3} (e), leads to a channel for exponential amplification in the open chain. The average particle distribution is then exponentially localized on the last unit cell. As $\gamma_1$ is increased, more and more particles dissipate into the environment preventing them from reaching the end of the chain. At some critical value $\gamma_1 = \gamma_c$, the dissipation is such that for a particle starting on the first site of a given unit cell, the non-reciprocity is perfectly nullified by the dissipation. In this case, the particles are uniformly distributed between the unit cells [\cref{fig:L3} (f)]---although they might not be uniformly distributed within each unit cell. It is at this point that there is a topological phase transition and the PBC spectral curve touches the origin. To find the critical value $\gamma_c$ we look for a zero energy state of the PBC chain.  Since the system is periodic with a unit cell of $L$ sites this reduces to finding the determinant of an $L\times L$ (dynamical) matrix and equating it to zero.  Alternatively, we can solve the following difference equations for the wavefunction $\psi_{n,j}^{(k)} = e^{ikn}\phi_j$ where $j$ represents the position within the unit cell, $n$ is the unit cell number related to the lattice momentum $k$.
\begin{align}
    i(t+\Delta)\phi_{j-1} -i(t-\Delta) \phi_{j+1} &= 0, \quad j\neq 1,L \label{eq:Lodd_dynamical_eq_2} \\
     i(t+\Delta)e^{-ik}\phi_L - i(t-\Delta) \phi_2 - \frac{i\gamma_c}{2}\phi_1 &= 0, \label{eq:Lodd_dynamical_eq_1} \\
      i(t+\Delta)\phi_{L-1} -i(t-\Delta) e^{ik} \phi_{1} &= 0.
\end{align}
Recalling that \mbox{$e^{2r} = \frac{t+\Delta}{t-\Delta}$}, the three equations above lead to the following equation:
\begin{equation}
    (t+\Delta)e^{-ik}e^{r(L-1)}-(t-\Delta)e^{ik}e^{-r(L-1)}={\gamma_c \over 2}.
\end{equation}
The requirement that $\gamma_c$ is real leads to $k=0,\pm\pi$ and since $\gamma_c$ is positive only the $k=0$ is possible.  This finally leads to the simple condition:

\begin{align}
    \label{eq:g1_criticalvalue}
    \frac{\gamma_c}{2} &= (t+\Delta)e^{r(L-1)} - (t-\Delta)e^{-r(L-1)}.
\end{align}
Since the critical value $\gamma_c$ given in \cref{eq:g1_criticalvalue} is valid for any odd length unit cell, it must also be valid for a unit cell of one site. Indeed, setting $L=1$, we retrieve the critical value of the uniform case found in \cref{sec:uniform}. 

\section{Conclusion}\label{sec:Conclusion}
We have shown that the phase-dependent topological amplification of a classical signal sent into one end of the open BKC is remarkably robust to on-site losses. It persists for arbitrarily large losses on every other site when the length of the chain is even, drastically exceeding conventional expectations. To understand why, we looked at uniform and non-uniform dissipation where the system is divided into $N/L$ unit cells of length $L$. Using the complex spectrum of the periodic dynamical matrix as well as the singular values and the susceptibility of the open dynamical matrix, we have found exact dissipation values at which a topological phase transition occurs for various unit cell lengths and bath coupling configurations. When the unit cell has two sites, we show that there is a critical value at which the difference in on-site dissipation induces a PBC energy band separation. Nonetheless, exponential amplification is observed as long as one PBC band winds around the origin in the complex plane. The system transitions out of the exponentially-amplifying topological phase when the geometric mean of the on-site dissipation of the two sites equals twice the nearest-neighbor drive: $\sqrt{\gamma_1\gamma_2}/2 = 2\Delta$. As such, the non-trivial topological phase of exponential amplification can be guaranteed by compensating an arbitrarily large dissipation on the first site by an arbitrarily small dissipation on the second. 
Looking at larger unit cells, we then separated our analysis between even and odd unit cell sizes, $L$. We showed that dissipation on the first site of unit cells of even length is never enough to induce a phase transition out of the topologically-amplifying phase. In contrast, it is enough for odd length unit cells and we find the critical value at which it happens. We explain this discrepancy via a non-Hermitian particle-hole symmetry and by looking at the limit of infinite loss.

We emphasize that our results follow directly from non-reciprocity, the defining feature of the Hatano-Nelson model. As such, we expect the main results to be adaptable to other bosonic systems possessing non-Hermitian dynamics.
All in all, our work demonstrates that the BKC is a promising quantum sensor---one that showcases strong signal amplification while being robust to disorder.

\begin{acknowledgments}
    We thank Alexander McDonald and Evgeny Moiseev for useful discussions. We acknowledge the support from NSERC, FRQNT, Qu\'ebec’s Minist\`ere de l’\'Economie, de l’Innovation et de l’ \'Energie (MEIE), Photonique Quantique Qu\'ebec (PQ2) and the Regroupement qu\'eb\'ecois sur les mat\'{e}riaux de pointe (RQMP).
\end{acknowledgments}

\appendix 

\section{Particle Distribution for Uniform Dissipation} \label{apdx:uniform}

To find an analytical expression of the particle distribution of the BKC subject to uniform dissipation, it is useful to consider the on-site squeezing transformation 
\begin{align}
    \hat{\mathcal{U}} &= \prod_{j=1}^N \exp\left[\frac{r(j-j_0)}{2}\left(\hat{a}_j \hat{a}_j - \hat{a}_j^\dagger \hat{a}_j^\dagger\right)\right]
\end{align}
where the squeezing parameter $r>0$ is defined by the ratio $e^{2r} \equiv \frac{t+\Delta}{t-\Delta}$ and $j_0$ is an arbitrary real number. Under this unitary transformation the BKC Hamiltonian of \cref{eq:BKC} becomes
\begin{align}
    \label{eq:BKC_tightbinding}
    \hat{\tilde{\mathcal{H}}} \equiv \hat{\mathcal{U}} \hat{\mathcal{H}} \hat{\mathcal{U}}^\dagger &= i\tilde{t}\sum_{j=1}^{N-1}\left(\hat{\tilde{a}}_{j+1}^\dagger\hat{\tilde{a}}_{j} -\hat{\tilde{a}}_{j}^\dagger\hat{\tilde{a}}_{j+1}\right)
\end{align}
where $\hat{\tilde{a}}_j = \cosh(rj)\hat{a}_j + \sinh(rj)\hat{a}_j^\dagger$ are the squeezed bosonic operators and $\tilde{t} \equiv \sqrt{t^2-\Delta^2} = t/\cosh(r)$ is an imaginary hopping parameter. The dissipationless BKC is thus unitarily-equivalent to a simple particle-conserving tight-binding model. Crucially, the quadrature susceptibilities in \cref{eq:chixx,eq:chipp} are given by
\begin{align}
    \chi^{xx}[j,\ell;\omega] &= e^{r(j-\ell)}\tilde{\chi}[j,\ell;\omega], \label{eq:quadrature_chi_to_tilde_xx}\\
     \chi^{pp}[j,\ell;\omega] &= e^{-r(j-\ell)}\tilde{\chi}[j,\ell;\omega], \label{eq:quadrature_chi_to_tilde_pp}
\end{align}
where the frequency space susceptibility $\tilde{\chi}$ of the tight-binding Hamiltonian of \cref{eq:BKC_tightbinding} of is known to be \cite{McDonald2020}
\begin{align}
    \label{eq:chi_free_BKC}
    \tilde{\chi}[j,\ell;\omega] &= i^{j-\ell+1}\frac{U_{\min(j,\ell)-1}\left(\frac{\omega}{2\tilde{t}}\right)U_{N-\max(j,\ell)}\left(\frac{\omega}{2\tilde{t}}\right)}{\tilde{t}U_N\left(\frac{\omega}{2\tilde{t}}\right)}
\end{align}
where $U_n$ is the Chebyshev polynomial of the second kind of degree $n$. Adding uniform dissipation $\gamma\geq 0$ on each site amounts to shifting the frequency $\omega\to \omega + i\gamma/2$ in $\tilde{\chi}$. The zero frequency (steady-state) response terms necessary to compute the average particle number are 
\begin{align}
    \tilde{\chi}[j,1;0] &= \frac{i^j U_{N-j}\left(i\frac{\gamma}{4\tilde{t}}\right)}{\tilde{t}U_N\left(i\frac{\gamma}{4\tilde{t}}\right)}.
\end{align}
The Chebyshev polynomial $U_n$ have the closed form 
\begin{align}
    U_n(x) &= \frac{\left(x+\sqrt{x^2-1}\right)^{n+1} - \left(x - \sqrt{x^2-1}\right)^{n+1}}{2\sqrt{x^2-1}}.
\end{align}
Setting $x = \frac{\gamma}{4\tilde{t}}$ and $y = x + \sqrt{x^2+1}$, we find 
\begin{align}
    U_n(ix) &= \frac{i^n}{2\sqrt{x^2+1}} \left[y^{n+1}-(-y)^{-(n+1)}\right].
\end{align}
As $y\geq 1$, the susceptibility terms for $N-j+1\gg 1$ can be written as
\begin{align}
    \tilde{\chi}[j,1;0] &= \frac{y^{N-j+1} - (-y)^{-(N-j+1)}}{\tilde{t}\left(y^{N+1} - (-y)^{-(N+1)}\right)} \simeq \frac{1}{\tilde{t}y^j}.
\end{align}
Since the average particle number is $\bar{n}_j \propto |\chi^{xx}[j,1;0]|^2$, we use \cref{eq:quadrature_chi_to_tilde_xx} to find that
\begin{align}
    \label{eq:ApdxA_photonnumber}
     \bar{n}_j &\propto \frac{1}{(t+\Delta)^2}\left(\frac{e^r}{y}\right)^{2j}.
\end{align}
The critical value $y_c = e^r$ is equivalent to $\gamma/2 = 2\Delta$ and corresponds to the case of uniform average particle distribution in the chain. When $y < e^r$, we have $\gamma/2 < 2\Delta$ and the particles localize exponentially on the last site of the chain. Lastly, $y > e^r$ corresponds to the situation of high dissipation $\gamma/2 > 2\Delta$ wherein particles are exponentially localized on the first site.  

\section{Localization Transition}\label{apdx:edge_transition}
Here we provide details on the eigenstate identified in \cref{sec:band_separation} that undergoes a transition in the edge on which it is localized. As mentioned in the main text, this state is an eigenstate of the open BKC whose unit cells of length $L>2$ are subject to loss $\gamma_1>0$ on their first site. 
We observe that there is a critical value $\gamma_1 = \zeta>0$ below which the eigenstate is localized on the edge $N$ with its energy inside the PBC spectral curve and above which it is localized on the other edge $j=1$ with its energy outside the PBC spectral curve. We emphasize that this eigenstate and its localization transition are not associated with the topological phase of amplification. To illustrate that such a state exists analytically, we focus on the case $L=3$ with semi-infinite boundary conditions (SIBC). We find $\zeta$ and show that the localization transition is equivalent to a change in the winding number of the PBC spectral curve about the energy of the SIBC eigenstate. Denote this eigenstate by $\ket{\psi} = \bigotimes_{j} z^j\ket{\phi}$ where $\ket{\phi} = (\phi_1,\phi_2,\phi_3)^T$ is the intra-unit-cell eigenstate and $z=z(\gamma_1)$ satisfies
\begin{align}
    \label{eq:z_regimes}
    |z(\gamma_1 < \zeta)| &> 1, \qquad |z(\gamma_1 > \zeta)| < 1
\end{align}
with $|z(\gamma_1 = \zeta)| = 1$ for some $\zeta>0$. To obtain an expression for $\zeta$, we solve the intra-unit-cell equation $M_{x}(z)\ket{\phi} = i\epsilon\ket{\phi}$
by imposing $|z(\gamma_1=\zeta)|=1$. Making the (numerically-informed) ansatz $\phi_3 = 0$, we find
\begin{align}
    \frac{\zeta}{2} &= \frac{(t+\Delta)^3 + (t-\Delta)^3}{(t+\Delta)(t-\Delta)}.
\end{align}
Comparing with the loss rate at which there is a topological phase transition $\gamma_c$ given in \cref{eq:g1_criticalvalue}, we find that $\zeta > \gamma_c$, since $t>\Delta>0$. In other words, the localization transition always happens after the topological phase transition. To show that there is a $z$ that satisfies \cref{eq:z_regimes}, we solve $M_x(z)\ket{\phi} = i\epsilon\ket{\phi}$ again but for $z$ in terms of $\gamma_1/\zeta$ to find 
\begin{align}
    \label{eq:z_value}
    z_\pm &= \frac{\gamma_1}{\zeta}\left(\frac{\pm\sqrt{(e^{6r} + 1)^2 - 4e^{6r}\frac{\zeta^2}{\gamma_1^2}} - (e^{6r}+1)}{2}\right)
\end{align}
where $e^{2r} = \frac{t+\Delta}{t-\Delta}$. While $|z_-|>1$ for all $\gamma_1$, we note that $z_+$ satisfies $z(\gamma_1=\zeta)=-1$ as well as \cref{eq:z_regimes}. Moreover, the energy of the mode $\ket{\psi}$ for $\gamma_1>0$ is 
\begin{align}
    i\epsilon &= -\frac{i\gamma_1}{4}\left(1 + \sqrt{1 - \frac{16(t^2-\Delta^2)}{\gamma_1^2}}\right),
\end{align}
which is $i\epsilon \simeq -\frac{i\gamma_1}{2}$ when $\gamma_1/2 \gg 2\sqrt{t^2-\Delta^2}$. Finally, the intra-unit-cell eigenstate $\ket{\phi}$ is 
\begin{align}
    \ket{\phi} = \frac{1}{\mathcal{N}}\begin{pmatrix}
        1 & \frac{t+\Delta}{\epsilon} & 0
    \end{pmatrix}^T.
\end{align}
We now show the equivalence between the localization transition of the SIBC eigenstate $\ket{\psi}$ and the change in winding number of the PBC spectral curve about the eigenstate's energy $i\epsilon$. Using \cref{eq:windingnum_def}, the argument principle gives that the winding number $\nu_{i\epsilon}(E)$ of the PBC curve $E=E(k)$ for $k\in(-\pi,\pi)$ about the point $i\epsilon\in \mathbb{C}$ is $\nu_{i\epsilon}(E) = Z-P$,
where $Z$ and $P$ are the zeroes and poles of $\tilde{E}(z) = E(z)-i\epsilon$ within the unit circle, respectively. While the zeroes correspond to when $E(z)=i\epsilon$, the number of poles of $\tilde{E}=\tilde{E}(z)$ corresponds to that of $E=E(z)$ since $i\epsilon$ is independent of $z$. From the characteristic equation $\det(M_x(z) - E)=0$, the function $E=E(z)$ possesses the single pole $z=0$ due to the matrix entry $-i(t-\Delta)z^{-1}$. Further, the number of $z\in\mathbb{C}$ with $|z|<1$ that satisfy $E(z) = i\epsilon$ can be obtained by solving the characteristic equation $\det(M_x(z)-i\epsilon)=0$. One then obtains $z=z(\gamma_1)$ given by \cref{eq:z_value}. From \cref{eq:z_regimes}, we know that $\gamma_1=\zeta$ separates a regime where $z_+$ lies outside of the unit circle to one where it lies inside. Therefore, the value $\gamma_1=\zeta$ separates two phases with different winding number: 
\begin{align}
    \nu_{i\epsilon}(E) &= \begin{dcases}
        -1 & \gamma_1 < \zeta, \\
        0 &\gamma_1 > \zeta.
    \end{dcases}
\end{align}
Finally, we argue that an eigenstate exhibiting such a localization transition cannot exist for unit cell sizes $L=1,2$. First, if $L=1$, then the system is completely translation-invariant for all $\gamma_1\geq0$. The OBC spectrum thus lies inside the area enclosed by the PBC spectral curve for all $\gamma_1\geq0$, in the sense that the winding number of the PBC curve about any OBC eigenenergy is always nonzero \cite{Okuma2020,Okuma2023}. For $L=2$, both the PBC curve $E(k)$ and the OBC spectrum are symmetric about the imaginary line $-\frac{i\gamma_1}{4}$. This symmetry stems from the system being parity-time symmetric, up to translation by $\frac{i\gamma_1}{4}$ (note that the system also has a similar symmetry for $L=1$, but not for $L>2$). Therefore, such a purely imaginary OBC eigenenergy with $\text{Im }E_- < \text{Im } E(k)$ for all $k\in(-\pi,\pi)$ would be accompanied by another solution with $\text{Im } E_+ > \text{Im } E(k)$, which would be dynamically unstable, see \cref{subsec:even_unit_cell}. Since increasing $\gamma_1\geq 0$ only creates more loss, it cannot spontaneously induce a dynamically unstable mode. 

\section{Susceptibility for the BKC with Two-Site Unit Cell} \label{apdx:L2}
Using the equation of motion technique, we derive a closed form for the susceptibility of the system used in \cref{subsec:even_unit_cell}. Similarly to \cref{apdx:uniform}, we start by deriving the susceptibility of a dissipative tight-binding model under OBC, here given by
\begin{align}
    \hat{\tilde{H}} &= i\tilde{t}\sum_{j=1}^{N-1} \left(\hat{\tilde{a}}_{j+1}^\dagger \hat{\tilde{a}}_j - \hat{\tilde{a}}_j^\dagger\hat{\tilde{a}}_{j+1}\right) +\sum_{j=1}^N -\frac{i\gamma_j}{2}\hat{\tilde{a}}_j^\dagger\hat{\tilde{a}}_j
\end{align}
where $\gamma_j=\gamma_1\geq 0$ when $j$ is odd and $\gamma_j=\gamma_2\geq 0$ when $j$ is even. Tthe retarded susceptibility is defined by
\begin{align}
    \tilde{\chi}(j,\ell;t) &= \Theta(t) \braket{[\hat{\tilde{a}}_j(t),\hat{\tilde{a}}^\dagger_\ell(0)]}
\end{align}
and so
\begin{equation}
    \begin{split}
        i\partial_t\tilde{\chi}(j,\ell;t) &= i\delta(t)\delta_{j,\ell} + i\tilde{t}\tilde{\chi}(j-1,\ell;t) \\
        &\quad- i\tilde{t}\tilde{\chi}(j+1,\ell;t) - \frac{i\gamma_j}{2}\tilde{\chi}(j,\ell;t).
    \end{split}
\end{equation}
Taking the Fourier transform of $\tilde{\chi}$ with respect to $t$ and denoting $\omega_j \equiv \omega + \frac{i\gamma_j}{2}$, 
\begin{equation}
    \begin{split}
        \omega_j\tilde{\chi}[j,\ell;\omega] &= i\delta_{j,\ell} + i\tilde{t}\tilde{\chi}[j-1,\ell;\omega] - i\tilde{t}\tilde{\chi}[j+1,\ell;\omega].
    \end{split}
\end{equation}
Up to some factor of $i$ and with the boundary conditions $\tilde{\chi}[N+1,\ell;\omega]=0$ and $\tilde{\chi}[0,\ell;\omega]=0$, the source-free recurrence relation is that of the Chebyshev polynomial of the second kind $U_j$, which satisfies $U_{j+1}(x) = 2x U_j(x) - U_{j-1}(x)$ as well as $U_0(x) = 1$ and $U_1(x) = 2x$. The solution is readily found to be 
\begin{align}
    \label{eq:L2_susceptibility}
    \tilde{\chi}[j,\ell;\omega] &= \frac{i^{j-\ell+1}U_{\min(j,\ell)-1}(\alpha)U_{N-\max(j,\ell)}(\alpha)}{\tilde{t}U_N(\alpha)}\rho_{j,\ell}
\end{align}
where $\alpha = \frac{\sqrt{\omega_1\omega_2}}{2\tilde{t}}$ and 
\begin{align}
    \rho_{j,\ell} &= \sqrt{\frac{\omega_{\max(j,\ell)-1}}{\omega_{\min(j,\ell)}}}.
\end{align}
Repeating the same procedure outlined in \cref{apdx:uniform} using \cref{eq:quadrature_chi_to_tilde_xx}, we find that the average particle number is uniformly distributed between the unit cells at the critical value 
\begin{align}
    \frac{\sqrt{\gamma_1\gamma_2}}{2} &= 2\Delta,
\end{align}
which corresponds precisely to the critical value separating the two topological phases that was found in \cref{eq:winding_number_L2}. On the one hand, when $\frac{\sqrt{\gamma_1\gamma_2}}{2}<2\Delta$, the particles are exponentially localized on the last site of the chain. On the other hand, $\frac{\sqrt{\gamma_1\gamma_2}}{2} > 2\Delta$ yields a particle distribution that is exponentially localized on the first site. 

\section{Line Gap Topology in Two-Site Unit Cell} 
\label{apdx:gainloss_balance}

\begin{figure}[t!]
     \centering
     \begin{tabular}{cc}
        \hspace{0.3cm}
        \begin{subfigure}[t]{0.32\textwidth}
            \centering
            \includegraphics[width=\textwidth]{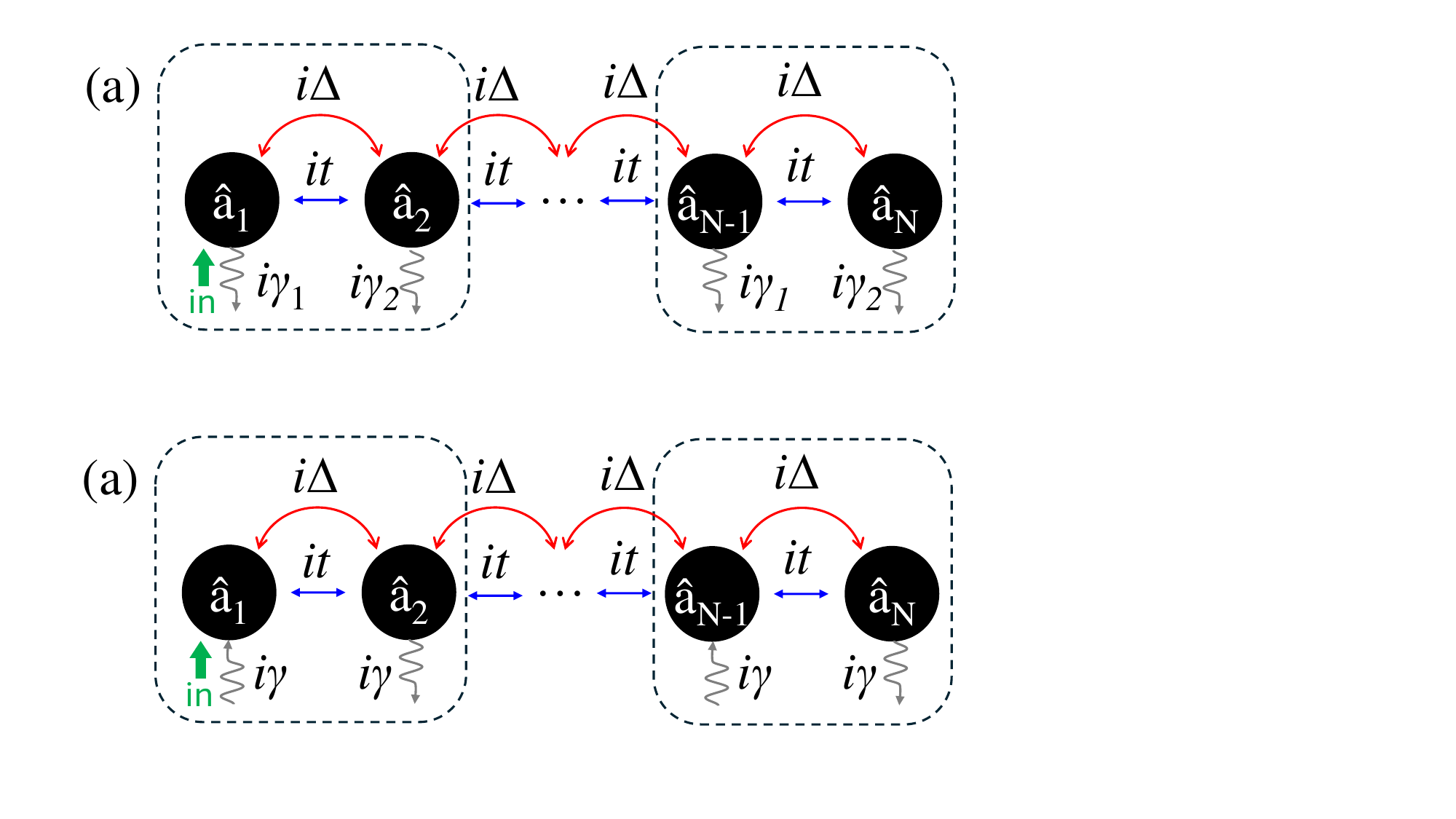}
        \end{subfigure} 
        \\
        \begin{tabular}[b]{ccc}
            \begin{subfigure}{0.154\textwidth}
            \centering
            \includegraphics[width=\textwidth]{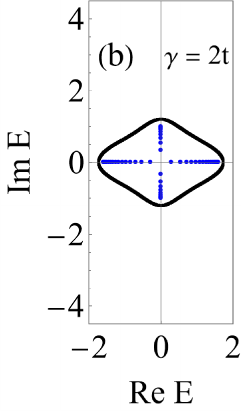}
            \end{subfigure}
            &
            \begin{subfigure}{0.124\textwidth}
                \centering
                \includegraphics[width=\textwidth]{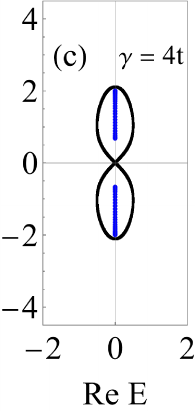}
            \end{subfigure} 
            &
            \begin{subfigure}{0.124\textwidth}
                \centering
                \includegraphics[width=\textwidth]{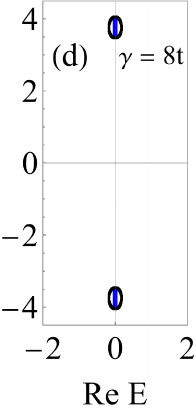}
            \end{subfigure}
        \end{tabular} 
        \\
        \begin{tabular}{cc}
             \begin{subfigure}[t]{0.21\textwidth}
            \centering
            \includegraphics[width=\textwidth]{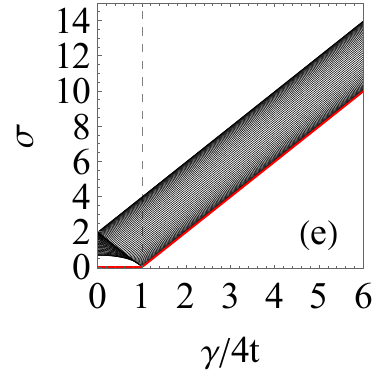}
        \end{subfigure}
        & 
        \begin{subfigure}[t]{0.238\textwidth}
            \centering
            \includegraphics[width=\textwidth]{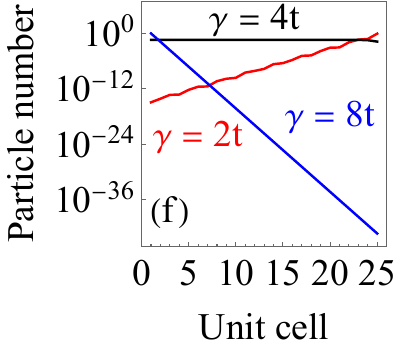}
        \end{subfigure}
        \end{tabular}
     \end{tabular}
    \caption{\justifying (a) Schematic, (b)--(d) spectra, (e) singular values and (f) normalized steady-state unit cell particle number distribution for the two-site unit cell BKC with 50 sites (25 unit cells) and balanced gain and loss. In (b)--(d), the PBC (black) and OBC (blue) spectrum of $M_x$ are shown for different coupling constants $\gamma$. There is a topological phase transition at $\gamma = 4t$. (e) Singular values of $M_x$ for the open BKC as a functon of $\gamma$. The critical dashed line $\gamma = 4t$ separates a regime wherein the smallest singular value (red) is exponentially small in system size from one in which it is not. In all plots $t=1$ and $\Delta=1/3$.}
    \label{fig:L2_balance}
\end{figure} 

Instead of having dissipation $-i\gamma_1/2$ and $-i\gamma_2/2$ on both sites of the two-site unit cell, suppose the first site experiences gain and the second site experiences loss. Balancing gain and loss, we have $\gamma_1 = -\gamma_2 = \gamma>0$. In this case, the full Hamiltonian made up of $N/2$ unit cells is parity-time ($PT$) symmetric; it commutes with the operator $PT$, where $P$ is the parity operator and $T$ is responsible for time reversal \cite{Bender98}. Adapting the analysis given in the beginning of \cref{subsec:even_unit_cell}, we find that the dynamical block $M_x(k)$ only has one connected energy level when $\gamma/2<2t$ and has two when $\gamma/2>2t$. Accordingly, the critical value $\gamma/2=2t$ corresponds to the emergence of an imaginary line gap in the PBC spectrum. At the same time, the susceptibility of the system \cref{eq:L2_susceptibility} in steady state has 
\begin{align}
    \tilde{\chi}[j,1;0] &= \frac{i^jU_{N-j}(\frac{\gamma}{4\tilde{t}})}{\tilde{t}U_N(\frac{\gamma}{4\tilde{t}})}\sqrt{(-1)^j}.
\end{align}
Setting $x = \frac{\gamma}{4\tilde{t}}$ and $y = x+\sqrt{x^2-1}$, we find 
\begin{align}
    U_n(x) &= \frac{y^{n+1} - y^{-(n+1)}}{2\sqrt{x^2-1}}.
\end{align}
Since $|y|\geq 1$, we can drop the negative exponent terms when $N-j\gg 1$ such that
\begin{align}
    \tilde{\chi}[j,1;0] &\simeq \frac{i^j\sqrt{(-1)^j}}{\tilde{t} y^j}.
\end{align}

Using \cref{eq:quadrature_chi_to_tilde_xx}, the particle number is again found to be given by \cref{eq:ApdxA_photonnumber}. However, since $y$ is different here, the critical value $y_c = e^r$ is equivalent to $\gamma/2 = 2t$. That is, when $\gamma/2<2t$, the PBC system has one energy level and the particle number distribution is exponentially localized on the last site of the open chain. When $\gamma/2>2t$, the PBC spectrum corresponds to two disconnected loops and the particle number distribution is localized on the first site. Lastly, the particle number is uniformly distributed in the open chain at $\gamma/2 = 2t$. Thus, $PT$-symmetry makes the opening of the line gap topological. See \cref{fig:L2_balance}.

\section{Susceptibility for the BKC with Arbitrary Odd Bath Coupling Constants}
\label{apdx:Neven_robustness_proof}
We prove that the BKC of even length $N$ exhibits exponential amplification independently of the values of odd bath coupling constants $\gamma_1,\gamma_3,\dots,\gamma_{N-1}\geq 0$, provided that $\gamma_2=\gamma_4=\dots=\gamma_N=0$. To this end, we let $\tilde{\chi}$ denote the susceptibility of the dissipationless BKC given in \cref{eq:chi_free_BKC}. Using Dyson's equation, we algebraically solve for the susceptibility $\tilde{\chi}_1$ of the system with dissipation on the first site, which corresponds to adding the term $(\tilde{V}_1)_{n,m} = -\frac{\gamma_1}{2}\delta_{n,1}\delta_{1,m}$ to the dynamical matrix. To incorporate dissipation on the third site, we add $(\tilde{V}_3)_{n,m} = -\frac{\gamma_3}{2}\delta_{n,3}\delta_{3,m}$ to the dynamical matrix and denote the associated susceptibility $\tilde{\chi}_3$. We repeat this procedure inductively until dissipation has been added on all odd sites, denoting 
by $\tilde{\chi}_{2j+1}$ the susceptibility of the system with $\tilde{V}_1,\tilde{V}_3,\dots,\tilde{V}_{2j+1}$. In the end, we prove by induction that $\tilde{\chi}_{N-1}[n,1;\omega=0] = \tilde{\chi}[n,1;\omega=0]$ for all $n=1,\dots,N$. That is, the system's steady-state response on site $n$ to a driving force on site 1 is independent of the fact that there is loss on the odd sites. We now proceed with the proof. First, note that the susceptibility of the dissipationless system given in \cref{eq:chi_free_BKC} has
\begin{align}
    \tilde{\chi}[n,1;\omega=0] = -\tilde{\chi}[1,n;\omega=0] = \begin{dcases}
        0 &\text{$n$ odd}, \\
        \tilde{t}^{-1} &\text{$n$ even}
    \end{dcases}
\end{align}
as well as $\tilde{\chi}[\ell,\ell';0]=0$ for any odd $1\leq \ell,\ell'\leq N$. Adding $\tilde{V}_1$ to the dynamical matrix, Dyson's equation yields 
\begin{align}
    \tilde{\chi}_1[n,m;\omega] &= \tilde{\chi}[n,m;\omega] - \frac{\frac{\gamma_1}{2}\tilde{\chi}[n,1;\omega]\tilde{\chi}[1,m;\omega]}{1+\frac{\gamma_1}{2}\tilde{\chi}[1,1;\omega]} 
\end{align}
and in steady state ($\omega=0$), we find
\begin{align}
    \tilde{\chi}_1[n,1;0] &= -\tilde{\chi}_1[1,n;0] = \tilde{\chi}[n,1;0]. 
\end{align}
Additionally, $\tilde{\chi}_1[\ell,\ell';0] = 0$ for any odd $1\leq \ell,\ell'\leq N$. Now, assume by induction that for some $j\geq 1$, 
\begin{align}
    \tilde{\chi}_{2j-1}[n,1;0] = -\tilde{\chi}_{2j-1}[1,n;0] = \tilde{\chi}[n,1;0]
\end{align}
as well as $\tilde{\chi}_{2j-1}[\ell,\ell';0]=0$ for any odd $1\leq \ell,\ell'\leq N$. Applying Dyson's equation for the next dissipative term $\tilde{V}_{2j+1}$, we readily obtain 
\begin{align}
    \tilde{\chi}_{2j+1}[n,1;0] = -\tilde{\chi}_{2j+1}[1,n;0] = \tilde{\chi}_{2j-1}[n,1;0]
\end{align}
and $\tilde{\chi}_{2j+1}[\ell,\ell';0]=0$ for any odd $1\leq \ell,\ell'\leq N$. By induction, the susceptibility $\tilde{\chi}_{N-1}$ of the even length BKC with arbitrary coupling constants $\gamma_1,\gamma_3,\dots,\gamma_{N-1}\geq 0$ on odd sites has $\tilde{\chi}_{N-1}[\ell,\ell';0]=0$ for any odd $1\leq \ell,\ell'\leq N$ as well as 
\begin{align}
    \tilde{\chi}_{N-1}[n,1;\omega=0] = \tilde{\chi}[n,1;\omega=0] = \begin{dcases}
        0 &\text{$n$ odd}, \\
        \tilde{t}^{-1} &\text{$n$ even}.
    \end{dcases}
\end{align}
Using \cref{eq:quadrature_chi_to_tilde_xx}, we conclude  that the average particle number distribution is exponentially localized on the last site $N$, showing that system is in a non-trivial topological phase for arbitrary values of $\gamma_1,\gamma_3\dots,\gamma_{N-1}\geq 0$.

\bibliography{references.bib}

\end{document}